\documentclass[a4paper, 12pt]{article}

\usepackage{tikz}
\usepackage{latexsym,amsmath,amsfonts,amssymb}
\usepackage{amsthm}
\usepackage{mathtools}
\usepackage[latin1]{inputenc}
\usepackage[american]{babel}
\usepackage{bbm}
\usepackage[nosort]{cite}
\usepackage[colorlinks=true, citecolor=blue, linkcolor=blue, linktocpage=true]{hyperref}

\renewcommand{\baselinestretch}{1.2}
\newcommand{\parfrac}[2]{\frac{\partial #1}{\partial #2}}

\newcommand{\wt}{\widetilde}
\newcommand{\wh}{\widehat}
\newcommand{\wb}{\overline}
\newcommand{\matht}[1]{\ensuremath{\boldsymbol{#1}}}

\newcommand{\ds}{\displaystyle}
\newcommand{\hiddensubsection}[1]{
    \stepcounter{subsection}
    \subsection*{\Alph{section}.\arabic{subsection}\hspace{1em}{#1}}
}

\newcommand{\ie}{\textit{i.e.}}

\numberwithin{equation}{section}

\newcommand{\nn}{\nonumber}
\newcommand{\mat}[1]{\begin{pmatrix} #1 \end{pmatrix}}

\newcommand{\be}{\begin{equation}} \newcommand{\ee}{\end{equation}}
\newcommand{\bea}{\begin{equation} \begin{aligned}} \newcommand{\eea}{\end{aligned} \end{equation}}
\newcommand{\ba}{\begin{array}} \newcommand{\ea}{\end{array}}

\newcommand{\cG}{\mathcal{G}}

\newcommand{\cI}{\mathcal{I}}
\newcommand{\cJ}{\mathcal{J}}

\newcommand{\cN}{\mathcal{N}}
\newcommand{\cO}{\mathcal{O}}

\newcommand{\cQ}{\mathcal{Q}}

\newcommand{\cS}{\mathcal{S}}

\newcommand{\cZ}{\mathcal{Z}}

\newcommand{\bC}{\mathbb{C}}

\newcommand{\bR}{\mathbb{R}}

\newcommand{\bZ}{\mathbb{Z}}

\newcommand{\fu}{\mathfrak{u}}

\newcommand{\unit}{\mathbbm{1}}

\newcommand{\fsu}{\mathfrak{su}}

\newcommand{\fso}{\mathfrak{so}}

\DeclareMathOperator{\Tr}{Tr}

\DeclareMathOperator{\re}{\mathbb{R}e}
\DeclareMathOperator{\im}{\mathbb{I}m}

\DeclareMathOperator{\diag}{diag}
\DeclareMathOperator{\Li}{Li}

\makeatletter
\def\blfootnote{\gdef\@thefnmark{}\@footnotetext}
\makeatother

\newcommand{\Abs}[1]{\left\vert#1\right\vert}

\begin{document}

\thispagestyle{empty}
\begin{flushright}
SISSA  56/2018/FISI
\end{flushright}
\vspace{25mm}  
\begin{center}
{\huge  Black holes in 4d $\cN=4$ Super-Yang-Mills 
}
\\[15mm]
{Francesco Benini$^{1,2,3}$, Paolo Milan$^{1,2}$}
 
\bigskip
{\it
$^1$ SISSA, Via Bonomea 265, 34136 Trieste, Italy \\[.5em]
$^2$ INFN, Sezione di Trieste, Via Valerio 2, 34127 Trieste, Italy \\[.5em]
$^3$ ICTP, Strada Costiera 11, 34151 Trieste, Italy \\[.5em]
}

{\tt fbenini@sissa.it, pmilan@sissa.it}

\bigskip
\bigskip

{\bf Abstract}\\[5mm]
{\parbox{14cm}{\hspace{5mm}

We resolve a long-standing question: does the four-dimensional $\cN=4$ $SU(N)$ Super-Yang-Mills theory on $S^3$ at large $N$ contain enough states to account for the entropy of rotating electrically-charged BPS black holes in AdS$_5$? Our answer is positive. We reconsider the large $N$ limit of the superconformal index, using the Bethe Ansatz formulation, and find an exponentially large contribution which exactly reproduces the Bekenstein-Hawking entropy of the black holes of Gutowski-Reall. Besides, the large $N$ limit exhibits a complicated structure, with many competing exponential contributions and Stokes lines, hinting at new physics.
}}
\end{center}

\newpage
\pagenumbering{arabic}
\setcounter{page}{1}
\setcounter{footnote}{0}
\renewcommand{\thefootnote}{\arabic{footnote}}

{\renewcommand{\baselinestretch}{1} \parskip=0pt
\setcounter{tocdepth}{2}
\tableofcontents}


\section{Introduction and results}
\label{sec: intro}

One of the fascinating aspects of black hole physics is its connection with the laws of thermodynamics. Of particular importance is the fact that black holes carry a macroscopic entropy \cite{Bekenstein:1972tm, Bekenstein:1973ur, Bekenstein:1974ax, Hawking:1974rv, Hawking:1974sw}, classically determined in terms of the horizon area. In the search for a theory of quantum gravity, explaining the microscopic origin of black hole thermodynamics is a fundamental but challenging test.

String theory is proposed to embed gravity in a consistent quantum system, hence it should in particular explain the black hole entropy in terms of a degeneracy of string states. This has been beautifully shown to be the case in the seminal paper \cite{Strominger:1996sh} by Strominger and Vafa, where the Bekenstein-Hawking entropy of a class of supersymmetric asymptotically-flat black holes was microscopically reproduced.

In the case of asymptotically-AdS black holes, the AdS/CFT duality \cite{Maldacena:1997re, Gubser:1998bc, Witten:1998qj} constitutes a natural and wonderful framework to study their properties at the quantum level. The duality provides a non-perturbative definition of quantum gravity, in terms of a conformal field theory (CFT) living at the boundary of AdS space. The problem of offering a microscopic account of the black hole entropy is rephrased into that of counting particular states in the dual CFT. However, despite the very favorable setup, this problem in four or more dimensions has remained unsolved for many years, and only recently a concrete example was successfully studied in \cite{Benini:2015eyy, Benini:2016rke}. There, the Bekenstein-Hawking entropy of a class of static dyonic BPS black holes in AdS$_4$ was holographically reproduced in the dual CFT$_3$, via supersymmetric localization \cite{Benini:2015noa, Benini:2016hjo, Closset:2016arn}. Since then, the matching has been extended to many other classes of magnetically-charged BPS black holes in various dimensions \cite{Hosseini:2016tor, Hosseini:2016cyf, Cabo-Bizet:2017jsl, Azzurli:2017kxo, Hosseini:2017fjo, Benini:2017oxt, Halmagyi:2017hmw, Bobev:2018uxk, Hosseini:2018uzp, Crichigno:2018adf, Suh:2018tul, Hosseini:2018usu, Suh:2018szn}, including the first quantum corrections \cite{Nian:2017hac, Liu:2017vll, Jeon:2017aif, Liu:2017vbl, Hosseini:2018qsx, Liu:2018bac}.

When moving to rotating, purely electric black holes, the situation becomes more complicated. Famously, the microstate counting for BPS black holes in AdS$_5$ has remained a long-standing open problem, which dates back to the work of \cite{Aharony:2003sx, Kinney:2005ej}. In this context, BPS black holes arise as rotating electrically-charged solutions of type IIB string theory on AdS$_5\times S^5$ \cite{Gutowski:2004ez, Gutowski:2004yv, Chong:2005da, Chong:2005hr, Kunduri:2006ek}. Their holographic description is in terms of 1/16 BPS states of the boundary 4d $\cN=4$ Super-Yang-Mills (SYM) theory on $S^3$, which can be counted (with sign) by the superconformal index \cite{Romelsberger:2005eg, Kinney:2005ej}. One would expect the contribution of the black hole microstates to the index  to dominate the large $N$ (\ie{} weak curvature) expansion. However, the large $N$ computation of the index performed in \cite{Kinney:2005ej} showed no rapid enough growth of the number of states, and thus it could not reproduce the entropy of the dual black holes. Additionally, that result was followed by several studies of BPS operators at weak coupling \cite{Berkooz:2006wc, Berkooz:2008gc, Janik:2007pm, Grant:2008sk, Chang:2013fba} in which no sign of high degeneracy of states was found.

Very recently, the issue received renewed attention leading towards a different conclusion. First, the authors of \cite{Cabo-Bizet:2018ehj} related the black hole entropy to the (complexified) regularized on-shell action of the gravitational black hole solutions, and then compared the latter with the $S^3 \times S^1$ supersymmetric partition function of the field theory, finding perfect agreement at leading order in large $N$. Second, the authors of \cite{Choi:2018hmj} analyzed the index in a double-scaling Cardy-like limit, finding quantitative evidence that the index does account for the entropy of large BPS black holes (whose size is much larger than the AdS radius). Third, in \cite{Choi:2018vbz} it was observed that, even at finite values of the fugacities, the index exhibits a deconfinement transition before the Hawking-Page transition related to the known AdS$_5$ black holes, pointing towards the existence of hairy black holes.

In this paper we offer a resolution of the issue by revisiting the counting of 1/16 BPS states in the boundary $\cN=4$ SYM theory at large $N$. We approach the problem by using a new expression for the superconformal index of the theory, derived in \cite{Closset:2017bse, Benini:2018mlo} and dubbed Bethe Ansatz (BA) formula, which allows for an easier analysis of the large $N$ limit. We find that the superconformal index, \ie{} the grand canonical partition function of 1/16 BPS states, does in fact grow very rapidly with $N$---as $e^{\cO(N^2)}$---for generic complex values of the fugacities. Although the BA formula of \cite{Benini:2018mlo} can handle the general case, this is technically difficult and in this paper we restrict to states and black holes with two equal angular momenta, as in \cite{Gutowski:2004yv}.

The BA formulation reveals that the large $N$ limit has a complicated structure. There are many exponentially large contributions, that somehow play the role of saddle points. As we vary the complex fugacities, those contributions compete and in different regions of the fugacity space, different contributions dominate. This gives rise to Stokes lines, separating different domains of analyticity of the limit. The presence of Stokes lines also resolves the apparent tension with the computation of \cite{Kinney:2005ej}, that was performed with real fugacities. We show that when the fugacities are taken to be real, all exponentially large contributions organize into competing pairs that can conceivably cancel against each other. The fact that for real fugacities the index suffers from strong and non-generic cancelations was already stressed in \cite{Choi:2018hmj, Choi:2018vbz}.

Our main result is to identify a particular exponential contribution, such that extracting from it the microcanonical degeneracy of states \emph{exactly} reproduces the Bekenstein-Hawking entropy of BPS black holes in AdS$_5$ (whose Legendre transform was obtained in \cite{Hosseini:2017mds}). This is in line with the double-scaling Cardy-like limit of \cite{Choi:2018hmj}. Along the way, we show that the very same $\cI$-extremization principle \cite{Benini:2015eyy, Benini:2016rke} found in AdS$_4$, is also at work in AdS$_5$ guaranteeing that the index captures the total number of single-center BPS black hole states.

At the same time, we step into many other exponentially large contributions: we expect them to describe very interesting new physics, that we urge to uncover. To that purpose, we study in greater detail the case of BPS black holes with equal charges and angular momenta \cite{Gutowski:2004ez}. We find that while for large black holes their entropy dominates the superconformal index, this is not so for smaller black holes. This seems to suggest%
\footnote{We are grateful to Shiraz Minwalla and Sameer Murthy for suggesting this possibility to us.}
that an instability, possibly towards hairy or multi-center black holes, might develop as the charges are decreased. Similar observations were made in \cite{Choi:2018hmj, Choi:2018vbz}. It would be extremely interesting if there were some connections with the recent works \cite{Bhattacharyya:2010yg, Markeviciute:2016ivy, Markeviciute:2018yal, Markeviciute:2018cqs}, and we leave this issue for future investigations.

The paper is organized as follows. In Section~\ref{sec: BHs} we review the charges and entropy of BPS black holes in AdS$_5$. In Section~\ref{sec: index} we present the BA formula for the superconformal index of $\cN=4$ SYM, and in Section~\ref{sec: large N} we compute its large $N$ limit. Sections~\ref{sec: stat interpretation} and \ref{sec: entropy from index} are devoted to extracting the black hole entropy from the index.


\section{BPS black holes in AdS\matht{_5}}
\label{sec: BHs}

In this paper we study the entropy of rotating charged BPS black holes in AdS$_5$ \cite{Gutowski:2004ez, Gutowski:2004yv, Chong:2005da, Chong:2005hr, Kunduri:2006ek} that can be embedded in type IIB string theory on AdS$_5\times S^5$ \cite{Cvetic:1999xp}. In order to set the stage, let us briefly review such gravitational solutions. The black holes are solutions to the equations of motion of type IIB supergravity that preserve one complex supercharge \cite{Gauntlett:2004cm}, thus being 1/16 BPS. The metric interpolates between the AdS$_5$ boundary and a fibration of AdS$_2$ on $S^3$ at the horizon. Moreover, the black holes carry three charges $Q_{1,2,3}$ for $U(1)^3\subset SO(6)$ acting on $S^5$, that appear as electric charges in AdS$_5$, and two angular momenta $J_{1,2}$ associated to the Cartan $U(1)^2\subset SO(4)$ (each Cartan generator acts on an $\bR^2$ plane inside $\bR^4$). The black hole mass is fixed by the linear BPS constraint
\be
\label{eq:mass}
M=g\Bigl( \Abs{J_1} + \Abs{J_2} + \Abs{Q_1} + \Abs{Q_2} + \Abs{Q_3} \Bigr) \;,
\ee
where $g=\ell_5^{-1}$ is the gauge coupling, determined in terms of the curvature radius $\ell_5$ of AdS$_5$ (whereas charges are dimensionless). It turns out that regular BPS black holes with no closed time-like curves only exist when the five charges satisfy certain non-linear constraints. The first constraint relies on the fact that one parameterizes the solutions by four real parameters $\mu_{1,2,3},\, \Xi$ \cite{Kunduri:2006ek}.%
\footnote{In \cite{Kunduri:2006ek} the authors use five real parameters $\mu_{1,2,3},\,a,\,b$  with $0 \leq a,b < g^{-1}$, however the black hole charges only depend on the combination $\Xi = \sqrt{(1-b^2g^2)/(1-a^2g^2)}$. The parameters $a,b$ are useful to write the full supergravity solutions. They are determined, in terms of $\mu_{1,2,3}$ and $\Xi$, by the extra relation
\be
\sqrt{(1-a^2g^2)(1-b^2g^2)} = \frac{2ab + 2g^{-1}(a+b) + 3g^{-2} }{ \mu_1 + \mu_2 + \mu_3 + 3g^{-2}} \;.
\ee
}
The second constraint is
\be
\label{eq:BHconstraint 2}
g^2\mu_{1,2,3} > \Xi - 1 \geq 0 \;.
\ee
Alternatively, one can have the same constraint with $\Xi$ substituted by $\Xi^{-1}$ which corresponds to exchanging $J_1 \leftrightarrow J_2$. The third constraint is
\be
\label{eq:BHconstraint 3}
S_\text{BH} \in \bR \;,
\ee
where the entropy $S_\text{BH}$ is defined in (\ref{eq:BHentropy}) below.

Charges and angular momenta of the black holes are completely determined by these four parameters $\mu_I, \Xi$ with $I=1,2,3$. Defining
\be
\gamma_1 = \sum_I \mu_I \;,\qquad \gamma_2 =  \sum_{I<J} \mu_I \mu_J \;,\qquad \gamma_3=\mu_1\mu_2\mu_3 \;,
\ee
the electric charges and angular momenta are
\bea
\label{eq:charges}
Q_I = \frac{\pi}{4 G_N} \left[ \frac{\mu_I}g + \frac{g}2 \left(\gamma_2-\frac{2\gamma_3}{\mu_I} \right) \right] \qquad
J_1 &= \frac{\pi}{4 G_N}\left[\frac{g \gamma_2}{2}+g^3\gamma_3+\frac{\cJ}{g^3} \biggl( \Xi -1 \biggr) \right] \\
J_2 &= \frac{\pi}{4 G_N}\left[\frac{g \gamma_2}{2}+g^3\gamma_3+\frac{\cJ}{g^3} \biggl( \frac1\Xi -1 \biggr) \right]
\eea
where $G_N$ is the five-dimensional Newton constant and
\be
\cJ=\prod_I \bigl( 1+g^2\mu_I \bigr) \;.
\ee
It is easy to see that one of the charges $Q_I$ can be zero or negative.%
\footnote{For instance, take $\mu_1$ that goes to zero with $\mu_{2,3}$ fixed, then $Q_1$ becomes negative. One may wonder whether the extra condition that the entropy be real could force the charges to be positive. This is not the case. For instance, setting $\mu_1 = \mu_2^2/ 3(1+\mu_2)$ and $\mu_3 = \mu_2$ as well as $\Xi=1$, one finds (up to constant factors and setting $g=1$) $Q_1 \sim - \mu_2^2/6 < 0$, $Q_2 = Q_3 \sim \mu_2 (\mu_2+2)/2>0$ and $S_\text{BH}^2 \sim \mu_2^4/12 > 0$.}
There are some combinations, though, that we can bound above zero, for instance:
\bea
\label{bounds on BH charges 1}
Q_1 + Q_2 + Q_3 &= \frac\pi{4G_N} \left[ \frac{\gamma_1}g + \frac{g \, \gamma_2}2 \right] > 0 \\
Q_I + Q_K &= \frac\pi{4G_N} \left[ \frac{\mu_I + \mu_K}g + g \, \mu_I \mu_K \right] > 0 \qquad\qquad\text{for } I \neq K \;.
\eea
In particular, at most one charge can be zero or negative. Setting $g=1$ for the sake of clarity, we also have
\bea
\label{bounds on BH charges 2}
Q_I + J_1 &= \frac{\pi}{4G_N} \biggl[ \Bigl( 1 + \mu_K \Bigr) \Bigl( 1 + \mu_L \Bigr) \Bigl( \mu_I + (1+\mu_I)(\Xi-1) \Bigr) \biggr] > 0 \\
Q_I + J_2 &= \frac{\pi}{4G_N} \biggl[ \Bigl( 1 + \mu_K \Bigr) \Bigl( 1 + \mu_L \Bigr) \biggl( \mu_I + (1+\mu_I)\Bigl( \frac1\Xi-1 \Bigr) \biggr) \biggr] >0
\eea
for $I\neq K \neq L \neq I$. The two inequalities follow from (\ref{eq:BHconstraint 2}).

The Bekenstein-Hawking entropy is proportional to the horizon area, and can be written as a function of the black hole charges \cite{Kim:2006he}:
\be
\label{eq:BHentropy}
S_\text{BH} = \frac{\text{Area}}{4G_N} = 2\pi \sqrt{Q_1Q_2+Q_1Q_3+Q_2Q_3 - \frac{\pi}{4G_Ng^3}\, \bigl( J_1+J_2 \bigr) \, } \;.
\ee
The constraint (\ref{eq:BHconstraint 3}) requires the quantity inside the radical to be positive. The BPS solutions have a regular well-defined event horizon only if the angular momenta are non-zero: in other words there is no static limit in gauged supergravity.

In this paper we will focus on the ``self-dual'' case $J_1 = J_2 \equiv J$ \cite{Gutowski:2004yv}. Since, in general, $\cJ>1$ and $\Xi\geq 1$, necessarily $\Xi=1$. The constraint (\ref{eq:BHconstraint 2}) simply becomes
\be
\mu_I > 0 \;.
\ee
The charges are
\be
\label{charges for J1=J2}
Q_I = \frac\pi{4G_N} \left[ \frac{\mu_I}g + \frac{g}2 \left( \gamma_2 - \frac{2\gamma_3}{\mu_I} \right) \right] \;,\qquad\qquad
J = \frac\pi{4G_N} \left[ \frac{g\gamma_2}2 + g^3\gamma_3 \right] > 0 \;.
\ee
The entropy is
\be
\label{eq:BHentropy J1=J2}
S_\text{BH} = \frac{2\pi^2}{4G_N} \sqrt{ \big( 1+g^2 \gamma_1 \big)\gamma_3 - \frac{g^2 \gamma_2^2}4 } = 2\pi \sqrt{ Q_1Q_2 + Q_1Q_3 + Q_2Q_3 - \frac\pi{4G_Ng^3} \, 2J \, } \;.
\ee
Once again, the constraint (\ref{eq:BHconstraint 3}) requires the quantity inside the radical to be positive.%
\footnote{We stress that the entropy is not automatically real. For instance, if we take $\mu_1$ that goes to zero with $\mu_{2,3}$ fixed, then the quantity inside the radical becomes negative. }


\section{The dual field theory and its index}
\label{sec: index}

A non-perturbative definition of type IIB string theory on AdS$_5\times S^5$ is in terms of its boundary dual: 4d $\cN=4$ SYM theory with $SU(N)$ gauge group \cite{Maldacena:1997re}, where
\be
\label{eq:dictionary}
N^2 = \frac{\pi \, {\ell_5}^3}{2 G_N}=\frac{\pi}{2G_N g^3} \;.
\ee
The weak curvature limit in gravity corresponds to the large $N$ and large 't~Hooft coupling limit in field theory. Up to the choice of gauge group, SYM is the unique four-dimensional Lagrangian CFT with maximal supersymmetry.
The field content, in $\cN=1$ notation, consists of a vector multiplet and three chiral multiplets $X,\,Y,\,Z$, all in the adjoint representation of the gauge group. Besides, there is a cubic superpotential $W=\Tr X[Y,Z]$. The R-symmetry is $SO(6)_R$: going to the Cartan $U(1)^3$, we choose a basis of generators $R_{1,2,3}$ each giving R-charge 2 to a single chiral multiplet and zero to the other two, in a symmetric way.

Considering the theory in radial quantization on $\bR\times S^3$, we are interested in the states that can be dual to the BPS black holes described in Section~\ref{sec: BHs}. These are 1/16 BPS states preserving one complex supercharge $\cQ$, and characterized by two angular momenta $J_{1,2}$ on $S^3$ and three R-charges for $U(1)^3 \subset SO(6)_R$. The angular momenta $J_{1,2}$ are semi-integer and each rotates an $\bR^2 \subset \bR^4$. Indicating with $J_\pm$ the spins under \mbox{$SU(2)_+ \times SU(2)_- \cong SO(4)$}, we set $J_{1,2} = J_+ \pm J_-$. With respect to the $\cN=1$ superconformal subalgebra (SCA) that contains $\cQ$, we describe the R-charges in terms of two flavor generators \mbox{$q_{1,2} = \frac{1}{2}(R_{1,2}-R_3)$} commuting with $\cQ$, and the R-charge $r=\frac{1}{3}(R_1+R_2+R_3)$. All fields in the theory have integer charges under $q_{1,2}$. The counting of BPS states is performed by the superconformal index \cite{Romelsberger:2005eg, Kinney:2005ej} defined by the trace
\be
\label{eq:traceindex}
\cI(p,q,v_1, v_2) = \Tr\,(-1)^F e^{-\beta\{\cQ,\cQ^\dagger\}} \, p^{J_1+\frac{r}{2}} \, q^{J_2+\frac{r}{2}} \, v_1^{q_1} \, v_2^{q_2} \;.
\ee
Here $p,q,v_a$ with $a=1,2$ are complex fugacities associated with the various quantum numbers, while the corresponding chemical potentials $\tau, \sigma, \xi_a$ are defined by
\be
p = e^{2\pi i\tau} \;,\qquad\qquad q = e^{2\pi i\sigma} \;,\qquad\qquad v_a = e^{2\pi i \xi_a} \;.
\ee
The fermion number is defined as $F = 2(J_+ + J_-) = 2J_1$. The index is well-defined for
\be
|p| \,, |q| < 1 \qquad\Leftrightarrow\qquad \im \tau \,,\, \im \sigma > 0 \;.
\ee
By standard arguments \cite{Witten:1982df}, $\cI$ only counts states annihilated by $\cQ$ and $\cQ^\dagger$ and is thus independent of $\beta$.

It will be convenient to redefine the flavor chemical potentials as
\be
\Delta_a = \xi_a + \frac{\tau+\sigma}3
\ee
and use
\be
y_a=e^{2\pi i \Delta_a} \;.
\ee
The index becomes%
\footnote{With respect to the notation in \cite{Kinney:2005ej}: $p = t^3x\big|_\text{there}$, $q=t^3/x\big|_\text{there}$, $y_1=t^2v\big|_\text{there}$, $y_2 = t^2w/v \big|_\text{there}$.}
\be
\label{eq:traceindex y}
\cI(p,q,y_1, y_2) = \Tr\,(-1)^F e^{-\beta\{\cQ,\cQ^\dagger\}} \, p^{J_1 + \frac12 R_3} \, q^{J_2 + \frac12 R_3} \, y_1^{q_1} \, y_2^{q_2} \;.
\ee
Notice that $J_1$, $J_2$, $\frac12 F$, $\frac12 R_3$ are all semi-integer and correlated according to
\be
\label{relation spin F R}
J_1 = J_2 = \frac F2 = \frac{R_3}2 \pmod{1} \;.
\ee
It is then manifest from (\ref{eq:traceindex y}) that the index is a single-valued function of the fugacities.

The index \eqref{eq:traceindex} admits an exact integral representation \cite{Romelsberger:2005eg, Kinney:2005ej, Dolan:2008qi}. In order to evaluate its large $N$ limit, though, we find more convenient to recast it in a different form, called Bethe Ansatz formula \cite{Closset:2017bse, Benini:2018mlo} (see also \cite{Closset:2018ghr} for a 3d analog, and \cite{Benini:2012ui, Doroud:2012xw, Benini:2013yva, Peelaers:2014ima, Benini:2016qnm} for similar Higgs branch localization formulas). Computing the large $N$ limit with this formula is still challenging, and in this paper we will restrict ourselves to the case of equal fugacities for the angular momenta:
\be
\tau = \sigma \qquad\Rightarrow\qquad p = q \;.
\ee
Hence, let us describe the Bethe Ansatz formula with this restriction \cite{Closset:2017bse},%
\footnote{In the notation of \cite{Benini:2018mlo}, that we will mostly follow, the restriction amounts to the case $a=b=1$.}
in the case of $\cN=4$ $SU(N)$ SYM. The superconformal index reads:
\be
\label{eq:BAindex}
\cI(q, y_1, y_2) = \kappa_{N}\sum_{\hat u\,\in\,\text{BAEs}} \cZ(\hat u;\Delta,\tau) \, H(\hat u;\Delta,\tau)^{-1} \;.
\ee
This is a finite sum over the solution set $\{\hat u\}$ to a system of transcendental equations, dubbed Bethe Ansatz Equations (BAEs), given by
\be
\label{eq:BAEs}
1 = Q_i(u; \Delta, \tau) = e^{2\pi i \left( \lambda + 3 \sum_j u_{ij} \right)} \prod_{j=1}^N \frac{ \theta_0(u_{ji} + \Delta_1; \tau) \, \theta_0(u_{ji} + \Delta_2; \tau) \, \theta_0(u_{ji} - \Delta_1 - \Delta_2; \tau) }{ \theta_0(u_{ij} + \Delta_1; \tau) \, \theta_0(u_{ij} + \Delta_2; \tau) \, \theta_0(u_{ij} - \Delta_1 - \Delta_2; \tau) }
\ee
for $i=1,\dots, N$ and where $u_{ij} = u_i - u_j$. We call $Q_i$ the BA operators. The unknowns are the ``complexified $SU(N)$ holonomies'' $u_i$ subject to the identifications
\be
u_i \,\sim\, u_i + 1 \,\sim\, u_i + \tau
\ee
meaning that each one lives on a torus of modular parameter $\tau$, and constrained by
\be
\label{SU constraint}
\sum_{i=1}^N u_i  = 0 \pmod{\bZ + \tau \bZ} \;,
\ee
as well as a ``Lagrange multiplier'' $\lambda$. The function $\theta_0$ is defined as
\be
\theta_0(u; \tau) = \prod_{k=0}^\infty \bigl( 1-zq^k \bigr) \bigl( 1- z^{-1} q^{k+1} \bigr) = (z;q)_\infty (q/z; q)_\infty \qquad\text{with } z=e^{2\pi i u} \,,\, q = e^{2\pi i \tau}
\ee
in terms of the $q$-Pochhammer symbol. Some of its properties are collected in Appendix~\ref{app: functions}. The prefactor in (\ref{eq:BAindex}) is
\be
\label{kappa_N}
\kappa_N =\frac{1}{N!}\Biggl( \frac{ (q;q)_\infty^2 \, \widetilde{\Gamma}(\Delta_1;\tau,\tau) \, \widetilde{\Gamma}(\Delta_2;\tau,\tau)}{\widetilde{\Gamma}(\Delta_1+\Delta_2;\tau,\tau)}\Biggr)^{N-1}
\ee
defined in terms of the elliptic gamma function \cite{Felder:1999}
\be
\label{eq:egamma}
\widetilde{\Gamma}(u;\tau,\sigma) = \Gamma\bigl(z=e^{2\pi i u};p=e^{2\pi i \tau}, q=e^{2\pi i\sigma}\bigr) = \prod_{m,n=0}^\infty \frac{1-p^{m+1}q^{n+1}/z}{1-p^mq^nz} \;.
\ee
The function $\cZ$ is
\be
\label{Z in BA formula}
\cZ(u; \Delta, \tau) = \prod_{i\neq j}^N \frac{ \wt\Gamma(u_{ij} + \Delta_1; \tau, \tau) \, \wt\Gamma(u_{ij} + \Delta_2; \tau, \tau) }{ \wt\Gamma(u_{ij} + \Delta_1 + \Delta_2; \tau, \tau) \, \wt\Gamma(u_{ij}; \tau, \tau)} \;.
\ee
Finally, the Jacobian $H$ is
\be
\label{Jacobian H}
H \Big|_\text{BAEs} = \det \left[ \frac1{2\pi i} \, \frac{\partial( Q_1, \dots, Q_N)}{\partial( u_1, \dots, u_{N-1}, \lambda)} \right]
\ee
when evaluated on the solutions to the BAEs. Notice that both $Q_i$, $\kappa_N$, $\cZ$ and $H$ are invariant under integer shifts of $\tau$, $\Delta_1$ and $\Delta_2$, implying that the superconformal index (\ref{eq:BAindex}) is a single-valued function of the fugacities.

Let us add some comments on how (\ref{eq:BAEs}) and (\ref{Jacobian H}) are obtained from the general formalism in \cite{Benini:2018mlo}. The maximal torus of $SU(N)$ is given by the matrices $\diag( z_1, \dots, z_{N-1}, z_N)$ with $\prod_{j=1}^N z_j = 1$ and, setting $z_j = e^{2\pi i u_j}$, is parameterized by $u_1, \dots, u_{N-1}$. For general gauge group $G$, the BA operators $Q_i$ have an index $i$ that runs over the Cartan subalgebra of $G$. Let us denote the BA operators of $SU(N)$ as $\wh Q_1, \dots, \wh Q_{N-1}$, then the BAEs are $\wh Q_j = 1$. The BA operators of $SU(N)$ can be written as $\wh Q_j = Q_j/ Q_N$ in terms of the BA operators $Q_1, \dots, Q_N$ of $U(N)$. Introducing a ``Lagrange multiplier'' $\lambda$, we can set $Q_N = e^{-2\pi i \lambda}$ and write the BAEs as $e^{2\pi i \lambda} Q_j = 1$ for $j=1, \dots, N$ (this includes the definition of $\lambda$). Absorbing $e^{2\pi i \lambda}$ into $Q_i$, we end up with (\ref{eq:BAEs}).

The Jacobian $H$ for $SU(N)$ is given by
\be
H = \det \biggl[ \frac1{2\pi i} \, \frac{\partial \wh Q_i}{\partial u_j} \biggr]_{i,j=1, \dots, N-1} \;.
\ee
When evaluated on the solutions to the BAEs, we have
\be
H \Big|_\text{BAEs} = \det \left[ \frac1{2\pi i} \, \frac{\partial (Q_i-Q_N)}{\partial u_j} \right]_{i,j=1, \dots, N-1} = (\ref{Jacobian H}) \;.
\ee
To see the last equality, one should  notice that $\partial Q_i/\partial \lambda \big|_\text{BAEs} = 2\pi i$.

The chemical potentials $u_j$ are defined modulo $1$, and the $SU(N)$ condition implies that they should satisfy $\sum_j u_j \in \bZ$. However, it is easy to check that the BAEs (\ref{eq:BAEs}) are invariant under shifts of one of the $u_j$'s by the periods of a complex torus of modular parameter $\tau$, namely $u_k \to u_k + n + m\tau$ for a fixed $k$. Hence the BAEs are well-defined on $N-1$ copies of the torus. Consistently, both $H$ and $\cZ$---when evaluated on the solutions to the BAEs---are invariant under shifts of $u_j$ by the periods of the torus (see \cite{Benini:2018mlo} for the general proof).

As one could suspect at this point, the BAEs (\ref{eq:BAEs}) are also invariant under modular transformations of the torus. To that purpose, it might be convenient to rewrite them in terms of the function $\theta(u;\tau) = e^{-\pi i u + \pi i \tau/6} \, \theta_0(u;\tau)$ that has simpler modular properties (see Appendix~\ref{app: functions}). When doing that, the term $\sum_j u_{ij}$ in the exponential in (\ref{eq:BAEs}) disappears. One easily shows that $Q_i$ are invariant under
\be
T: \left\{ \begin{aligned} \tau &\mapsto \tau+1 \\  u &\mapsto u \end{aligned} \right. \qquad\qquad S: \left\{ \begin{aligned} \tau &\mapsto - \frac1\tau \\ u &\mapsto \frac u\tau \end{aligned} \right.\qquad\qquad C: \left\{ \begin{aligned} \tau &\mapsto \tau \\ u &\mapsto -u \end{aligned} \right.
\ee
thus showing invariance under the full group $SL(2,\bZ)$. On the other hand, the summand $\kappa_N \cZ H^{-1}$ in (\ref{eq:BAindex}) is not invariant under modular transformations of $\tau$: this is not a symmetry of the superconformal index.

\subsection{Exact solutions to the BAEs}
\label{sec: exact solutions}

When evaluating the BA formula (\ref{eq:BAindex}), the hardest task is to solve the BAEs (\ref{eq:BAEs}). The very same equations appear in the $T^2 \times S^2$ topologically twisted index \cite{Benini:2015noa}, and one exact solution was found in \cite{Hosseini:2016cyf, Hong:2018viz}:
\be
\label{BA solution basic}
u_{ij} = \frac\tau N \, (j-i) \;,\qquad\qquad u_j = \frac{\tau \, (N-j)}N + \bar u \;,\qquad\qquad \lambda = \frac{N-1}2 \;. 
\ee
Here $\bar u$ is a suitable constant that solves the $SU(N)$ constraint (\ref{SU constraint}); since all expressions depend solely on $u_{ij}$, we will not specify that constant.
Notice that the solution does not depend on the chemical potentials $\Delta_a$. To prove that it is a solution, we compute
\begin{multline}
\prod_{j=1}^N \frac{\theta_0(u_{ji} + \Delta)}{\theta_0( u_{ij} + \Delta)} = \frac{\prod_{k=0}^{i-1} \theta_0 \bigl( \frac\tau N k + \Delta \bigr) \times \prod_{k=i-N}^{-1} \theta_0 \bigl( \frac\tau N k + \Delta \bigr) }{ \prod_{k=0}^{N-i} \theta_0 \bigl( \frac\tau N k + \Delta \bigr) \times \prod_{k=1-i}^{-1} \theta_0 \bigl( \frac\tau N k + \Delta \bigr) } = {} \\
= \frac{ \prod_{k=0}^{N-1} \theta_0 \bigl( \frac \tau N k + \Delta \bigr) \times \prod_{k=i-N}^{-1} \bigl( -q^{k/N} y \bigr) }{ \prod_{k=1}^{N-1} \theta_0 \bigl( \frac \tau N k + \Delta \bigr) \times \prod_{k=1-i}^{-1} \bigl( -q^{k/N} y \bigr) } = (-1)^{N-1} \, y^{N-2i+1} \, q^{i-\frac{N+1}2} \;.
\end{multline}
To go to the second line we used the periodicity relations (\ref{theta0 periodicities}). Taking the product over $\Delta = \{\Delta_1, \Delta_2, - \Delta_1 - \Delta_2\}$ we precisely reproduce the inverse of the prefactor of (\ref{eq:BAEs}), for every $i$.
Furthermore, notice that the shift $\bar u \to \bar u + \frac1N$ generates a new inequivalent solution that solves the $SU(N)$ constraint. Repeating the shift $N$ times, because of the torus periodicities, we go back to the original solution. Therefore, (\ref{BA solution basic}) actually represents $N$ inequivalent solutions.

Because the BAEs are modular invariant, we could transform $\tau$ to $\tau' = (a\tau + b)/(c\tau + d)$, then write the solution $u_{ij}' = \tau'(j-i)/N$, and finally go back to $\tau = (d\tau'-b)/(a-c\tau')$. This gives, for any $a,b\in \bZ$ with $\gcd(a,b)=1$, an $SL(2,\bZ)$-transformed solution
\be
\label{BA solutions SL transformed}
u_{ij} = \frac{ a\tau + b}N \, (j-i) \;.
\ee
However, one should only keep the solutions that are not equivalent---either because of periodicities on the torus or because of Weyl group transformations.

On the other hand, a larger class of inequivalent solutions was found in \cite{Hong:2018viz} (we do not know if this is the full set or other solutions exist). For given $N$, every choice of three non-negative integers $\{m,n,r\}$ that decompose $N = m \cdot n$ and with $0 \leq r < n$ leads to an exact solution
\be
\label{BA solutions Hong-Liu}
u_{\hat\jmath \hat k} = \frac{\hat\jmath}m + \frac{\hat k}n \left( \tau + \frac rm \right) + \bar u
\ee
where $\hat\jmath = 0, \dots, m-1$ and $\hat k = 0, \dots, n-1$ are an alternative parameterization of the index $j = 0, \dots, N-1$. As we show below, the first class is contained into the second class. Once again, (\ref{BA solutions Hong-Liu}) actually represents $N$ inequivalent solutions because of the possibility of shifting $\bar u$.

The solutions (\ref{BA solutions Hong-Liu}) organize into orbits of $PSL(2,\bZ)$ with the following action:
\be
T: \{m,n,r\} \mapsto \{m , n, r+m\} \;,\qquad S: \{m,n,r\} \mapsto \left\{ \gcd(n,r) \,,\, \frac{m\,n}{\gcd(n,r)} \,,\, \frac{m(n-r)}{\gcd(n,r)} \right\}
\ee
where the last entry of $\{m',n',r'\}$ is understood mod $n'$. One can check that $S^2 = \unit$. If $\{m,n,r\}$ have a common divisor, then one can see that also $\{m',n',r'\}$ have that common divisor, and since $T,S$ are invertible, it follows that $\gcd(m,n,r)\equiv d$ is an invariant along $PSL(2,\bZ)$ orbits.

We can prove that if $\{m,n,r\}$ have $\gcd(m,n,r)=1$, then they are in the orbit of $\{1,mn,0\}$, \ie{} there exists a $PSL(2,\bZ)$ transformation that maps them to $\{1,mn,0\}$. Indeed, let $\tilde r = \gcd(m,r)$. We can perform a number of $T$ transformations to reach $\{m,n, \tilde r\}$. Necessarily $\gcd(n, \tilde r)=1$, therefore an $S$ transformation gives $\{1, mn, m(n-\tilde r)\}$. Now a number of $T$ transformations gives $\{1, mn, 0\}$. On the other hand, we observe that if $\gcd(m,n,r)=d>1$, then the orbit under $PSL(2,\bZ)$ is in one-to-one correspondence with the one of $\{m/d, n/d, r/d\}$, which is generated by $\{1, mn/d^2, 0\}$. This shows that the number of orbits is equal to the number of divisors $d^2$ of $N$ which are also squares. Each orbit is generated by $\{d, N/d, 0\}$, and is in one-to-one correspondence with the orbit generated by $\{1, N/d^2,0\}$, which we can regard as the ``canonical form''.

At this point we recognize that the set of inequivalent solutions in the first class (\ref{BA solutions SL transformed}) (neglecting shifts of $\bar u$) is precisely the $PSL(2,\bZ)$ orbit with $\gcd(m,n,r)=1$ in the second class (\ref{BA solutions Hong-Liu}). Indeed, start with a solution of type (\ref{BA solutions SL transformed}) for some $N$ and some coprime integers $a,b$. Let $m = \gcd(a,N)$ and $n = N/m$. We can write the solution as
\be
u_j = - \frac{(a/m)\, j}n \tau - \frac{b\, j}N + \bar u \pmod{\bZ + \tau\bZ} \;.
\ee
We can identify $\hat k = (a/m) j \mod n$. Since $(a/m)$ and $n$ are coprime, as $j$ runs from $0$ to $n-1$, $\hat k$ takes all values in the same range once. Moreover there exists $s = (a/m)^{-1} \mod n$, such that $j = s\hat k \mod n$. In other words, $(a/m)$ is invertible mod $n$ and its inverse $s$ is coprime with $n$. We can write
\be
j = s\hat k + n \hat\jmath
\ee
and as $j$ runs from $0$ to $N-1$, $\hat\jmath$ covers a range of length $m$. Substituting the expression for $j$ we obtain
\be
u_j = - \frac bm \hat\jmath - \frac{\hat k}n \left( \tau + \frac{bs}{m} \right) + \bar u  \pmod{\bZ + \tau\bZ} \;.
\ee
Notice that $\gcd(b,m) = 1$. Indeed, suppose that $b$ and $m$ have a common factor, then this must also be a factor of $a$, which is a contradiction. Therefore we have the equality of sets \mbox{$\{ b \hat\jmath \mod m\} = \{\hat\jmath \mod m\}$}. Finally, we set $r = bs \mod n$ and we reproduce the expression in (\ref{BA solutions Hong-Liu}). The values $\{m,n,r\}$ obtained this way have $\gcd(m,n,r)=1$. Indeed, suppose they have a common factor, then this must also be a factor of $a$ but not of $(a/m)$, and thus it must also be a factor of $b$, which is a contradiction.

On the contrary, start with a solution $\{m,n,r\}$ of type (\ref{BA solutions Hong-Liu}) with $\gcd(m,n,r)=1$. It is easy to see, by repeating the procedure, that it is equivalent to a solution of type (\ref{BA solutions SL transformed}) with $a=m$ and $b=r$ (which imply $s=1$).

\section{The large \matht{N} limit}
\label{sec: large N}

In this Section we take the large $N$ limit of the BA formula (\ref{eq:BAindex}) for the superconformal index. The first part of the Section is technical, and the uninterested reader could directly jump to Section~\ref{sec: final result} where the final result is presented.

In the related context of the $T^2 \times S^2$ topologically twisted index \cite{Closset:2013sxa, Benini:2015noa}, it was shown in \cite{Hosseini:2016cyf} that the basic solution (\ref{BA solution basic}) leads to the dominant contribution in the high temperature limit. Assuming that such a solution gives an important contribution in our setup as well, we will start evaluating its large $N$ limit. We will find that it scales as $e^{\cO(N^2)}$, therefore in the following we will systematically neglect any factor whose logarithm is subleading with respect to $\cO(N^2)$. We will also find that the solution (\ref{BA solution basic}) is not necessarily dominant in our setup, rather other solutions can compete, and we will thus have to include the contributions of some of the solutions (\ref{BA solutions SL transformed}).

First of all, consider the prefactor $\kappa_N$ in (\ref{eq:BAindex}) and the multiplicity of the BA solutions, whose contribution does not depend on the particular solution. Each BA solution (\ref{BA solutions Hong-Liu}) has multiplicity $N \cdot N!$, where the first factor comes from shifts of $\bar u$ while the second factor from the Weyl group action. Thus, from (\ref{kappa_N}), we find
\be
N \cdot N! \cdot \kappa_N = e^{\cO(N)} \;.
\ee
This contribution can be neglected at leading order.

\subsection{Contribution of the basic solution}
\label{sec: contrib basic sol}

Here we consider only the contribution of the basic solution (\ref{BA solution basic}) to the sum in (\ref{eq:BAindex}).

\paragraph{The Jacobian.} We use the expression in (\ref{Jacobian H}). The derivative of $Q_i$ with respect to $u_j$ can be computed and it gives:
\be
\frac{\partial \log Q_i(u;\Delta,\tau)}{\partial u_j} = \sum_{k=1}^N \partial_{u_j} u_{ik} \left( 6\pi i + \sum_{\Delta \in \{\Delta_1, \Delta_2, -\Delta_1 - \Delta_2\}} \frac{\cG'(u_{ik};\Delta,\tau) }{ \cG(u_{ik};\Delta,\tau)} \right) \;,
\ee
with
\be
\cG(u;\Delta,\tau) = \frac{\theta_0(-u+\Delta;\tau)}{\theta_0(u+\Delta;\tau)}
\ee
and $\partial_{u_j} u_{ik} = \delta_{ij}-\delta_{kj} - \delta_{iN} + \delta_{kN}$. This relation holds because we take $u_1, \dots, u_{N-1}$ as the independent variables, and fix $u_N$ using (\ref{SU constraint}). Substituting we get
\begin{multline}
\label{eq:dQ/du}
\frac{\partial \log Q_i(u;\Delta,\tau)}{\partial u_j} = (\delta_{ij}-\delta_{iN}) \left( 6\pi i N + \sum_{k=1}^N \sum_\Delta \frac{\cG'(u_{ik} ;\Delta,\tau)}{\cG(u_{ik} ;\Delta,\tau)} \right) + {} \\
+ \sum_\Delta \left( \frac{\cG'(u_{iN};\Delta,\tau)}{\cG(u_{iN};\Delta,\tau)} - \frac{\cG'(u_{ij};\Delta,\tau)}{\cG(u_{ij};\Delta,\tau)} \right)
\end{multline}
where $\Delta$ is summed over $\{\Delta_1, \Delta_2, -\Delta_1 - \Delta_2\}$.

When we evaluate this expression on $u_{ij} = \tau(j-i)/N$, we notice that---for generic values of $\Delta_a$---the terms in the second line are of order $\cO(1)$. Indeed, the distribution of points $u_{ij}$ generically does not hit any zeros or poles of $\cG$. Retaining only the terms in the first line, the Jacobian reads
\be
H = \det\mat{ A_1 & \cO(1) & \cdots & \cO(1) & 1 \\ \cO(1) & A_2 & & \vdots & 1 \\ \vdots & & \ddots & \vdots & \vdots \\ \cO(1) & \cdots & \cdots & A_{N-1} & 1 \\ -A_N & -A_N & \cdots & -A_N & 1}
\ee
where the diagonal entries are
\be
\label{eq:A_i}
A_i = 3 N + \frac1{2\pi i} \sum_{k=1}^N \sum_\Delta \frac{\cG'(u_{ik} ;\Delta,\tau)}{\cG(u_{ik} ;\Delta,\tau)} \;.
\ee
Let us estimate the behavior of $A_i$ with $N$. By the same argument as above, $A_i$ contains the sum of $N$ elements of order $\cO(1)$ and thus it scales like $\cO(N)$ (or smaller). The determinant can be computed at leading order and it gives
\be
\label{eq:J}
H = \sum_{k=1}^N \prod_{j\, (\neq k)\, =1}^N A_j + \text{subleading} \;.
\ee
This scales as $\cO(N^N)$, therefore $\log H = \cO(N\log N)$ and can be neglected.

\paragraph{The functions \matht{\wt\Gamma}.} The dominant contribution comes from the function $\cZ$ defined in (\ref{Z in BA formula}). To evaluate it, let us analyze $\sum_{i\neq j}^N \log \wt\Gamma(u_{ij} + \Delta; \tau,\tau)$ with $\Delta \in \{\Delta_1,\Delta_2, \Delta_1 + \Delta_2\}$ separately. Making use of the relation (\ref{degenerate transformation Gamma}) proven in \cite{Felder:1999}, we write
\be
\label{modular expansion Gamma tilde}
\wt\Gamma(u_{ij} + \Delta; \tau, \tau) = \frac{ \rule[-.5em]{0pt}{0em} e^{\ds -\pi i \cQ \left( u_{ij}+ \Delta; \tau, \tau \right)} }{ \rule{0pt}{1.7em} \theta_0 \Bigl( \dfrac{u_{ij} + \Delta}\tau; - \dfrac1\tau \Bigr)} \; \prod_{k=0}^\infty \frac{ \rule[-1em]{0pt}{0em} \psi \Bigl( \dfrac{ k + 1 + u_{ij} + \Delta }\tau \Bigr) }{ \rule{0pt}{1.7em} \psi \Bigl( \dfrac{ k - u_{ij} - \Delta }\tau \Bigr) } \;.
\ee
The function $\psi(t)$ is defined in (\ref{def function psi}), while
\be
\label{def Q}
\cQ(u; \tau,\sigma) = \frac{u^3}{3\tau\sigma} - \frac{\tau + \sigma-1}{2\tau\sigma} u^2 + \frac{(\tau + \sigma)^2 + \tau\sigma - 3(\tau + \sigma) +1}{6\tau \sigma} u + \frac{(\tau + \sigma -1)(\tau + \sigma - \tau\sigma)}{12\tau\sigma}
\ee
is a cubic polynomial in $u$.

To make progress, we perform a series expansion of $\log\theta_0$ and $\log\psi$, evaluate this expansion on the basic solution for $u_{ij}$ in (\ref{BA solution basic}), and perform the sum $\sum_{i\neq j}^N$. We define the modular transformed variables
\be
\tilde z = e^{2\pi i u/\tau} \;,\qquad\qquad \tilde y = e^{2\pi i y / \tau} \;,\qquad\qquad \tilde q = e^{-2\pi i /\tau} \;.
\ee
We have
\bea
\label{log theta0 tau=sigma}
& \sum_{i\neq j}^N \log\, \theta_0\left( \frac{u_{ij} + \Delta}\tau; - \frac1\tau \right) = \sum_{n=0}^\infty \sum_{i\neq j}^N \log \left[ \left( 1 - \frac{\tilde z_i}{\tilde z_j} \, \tilde y \, \tilde q^n \right) \left( 1 - \frac{\tilde z_j}{\tilde z_i} \, \tilde y^{-1} \, \tilde q^{n+1} \right) \right] \\
&\qquad = - \sum_{\ell=1}^\infty \sum_{n=0}^\infty \sum_{i\neq j}^N \frac1\ell \left[ \left( \frac{\tilde z_i}{\tilde z_j} \, \tilde y \, \tilde q^n \right)^\ell + \left( \frac{\tilde z_j}{\tilde z_i} \, \tilde y^{-1} \, \tilde q^{n+1} \right)^\ell \right] \\
&\qquad = - \sum_{\ell=1}^\infty \sum_{n=0}^\infty \frac1\ell \Bigl[ A_\ell \; \tilde y^\ell \, \tilde q^{n\ell} + A_\ell \; \tilde y^{-\ell} \, \tilde q^{(n+1)\ell} \Bigr] = - \sum_{\ell=1}^\infty \frac1\ell \, A_\ell \, \frac{\tilde y^\ell + \tilde y^{-\ell} \tilde q^\ell}{ 1- \tilde q^\ell}
\eea
where we introduced $A_\ell$ which denotes the following sum over $ij$:
\be
A_\ell \,\equiv\, \sum_{i\neq j}^N \left( \frac{ \tilde z_i}{\tilde z_j} \right)^\ell = \sum_{i\neq j}^N e^{2\pi i (j-i)\ell/N} 
= \begin{cases} N^2 - N &\text{for } \ell=0 \mod N \\ -N &\text{for } \ell\neq 0 \mod N \;. \end{cases}
\ee
The series can be resummed to $N \log \left[ \theta_0\left( \frac{N\Delta}\tau; - \frac N\tau \right) / \theta_0\left( \frac\Delta\tau; - \frac1\tau \right) \right]$, however we do not need that. We collect the terms into two groups:
\be
\eqref{log theta0 tau=sigma} = N \sum_{\ell=1}^\infty \frac1\ell \, \frac{ \tilde y^\ell + \tilde y^{-\ell} \, \tilde q^\ell}{1-\tilde q^\ell} - N \sum_{j=1}^N \frac1j \ \frac{ \tilde y^{Nj} + \tilde y^{-Nj} \, \tilde q^{Nj} }{1-\tilde q^{Nj}}
\ee
where the second term comes from $\ell=Nj$. For $|\tilde q| < |\tilde y| < 1$, namely for
\be
\label{regime of convergence}
\im\left( - \frac1\tau \right)  > \im\left( \frac\Delta\tau \right)  > 0 \;,
\ee
the series converges. The second term is suppressed at large $N$, whereas the first term is of order $\cO(N)$ and can be neglected.

We then perform a similar analysis of $\log\psi$, using the series expansions of the functions $\log$ and $\Li_2$. We find
\begin{align}
\label{log psi tau=sigma}
& \sum_{i\neq j}^N \sum_{k=0}^\infty \log \frac{ \psi \bigl( \frac{ k+1 + u_{ij} + \Delta}\tau \bigr) }{ \psi \bigl( \frac{k - u_{ij} - \Delta }\tau \bigr) } = \sum_{i\neq j}^N \sum_{k=0}^\infty \sum_{\ell=1}^\infty \left[ - \frac1\ell \left( \frac{k+1+\Delta}\tau \tilde y^{-\ell} \, \tilde q^{\ell} - \frac{k-\Delta}\tau \tilde y^\ell \right) + \right. \\
&\hspace{5.5cm} \left. - \frac1\ell \, \frac{u_{ij}}\tau \bigl( \tilde y^\ell + \tilde y^{-\ell} \, \tilde q^\ell \bigr) + \frac1{2\pi i} \, \frac1{\ell^2} \bigl( \tilde y^\ell - \tilde y^{-\ell} \, \tilde q^\ell \bigr) \right] \left( \frac{\tilde z_i}{\tilde z_j} \, \tilde q^k \right)^\ell \nn \\
&\qquad = \sum_{k=0}^\infty \sum_{\ell=1}^\infty \left[ - \frac{A_\ell}\ell \left( \frac{k+1+\Delta}\tau \, \tilde y^{-\ell} \, \tilde q^{\ell} - \frac{k-\Delta}\tau \, \tilde y^\ell \right) \tilde q^{k\ell} + \frac1{2\pi i} \, \frac{A_\ell}{\ell^2} \bigl( \tilde y^\ell - \tilde y^{-\ell} \, \tilde q^\ell \bigr) \tilde q^{k\ell} \right] \nn
\end{align}
where we used that the following sum vanishes:
\be
B_\ell \equiv \sum_{i\neq j}^N \tilde u_{ij} \left( \frac{\tilde z_i}{\tilde z_j} \right)^\ell = \frac1N \sum_{i \neq j}^N (j-i) \, e^{2\pi i (j-i)\ell/N} = 0 \;.
\ee
Once again, the expression can be resummed by breaking the sum into two groups (corresponding to generic $\ell$ and $\ell = Nj$):
\be
\eqref{log psi tau=sigma} = \sum_{k=0}^\infty \left[ - N \log \frac{ \psi \bigl( \frac{k+1+\Delta}\tau \bigr) }{ \psi \bigl( \frac{k-\Delta}\tau \bigr) } + \log \frac{ \psi \bigl( \frac{ N(k+1-\Delta)}\tau \bigr) }{ \psi \bigl( \frac{ N(k-\Delta) }\tau \bigr) } \right] \;.
\ee
The first term (that comes from setting $A_\ell \to N$) is of order $\cO(N)$ and can be neglected. The second term is
\be
\sum_{k=0}^\infty \sum_{j=1}^\infty \left[ - \frac Nj \left( \frac{k+1+\Delta}\tau (\tilde q/ \tilde y)^{Nj} - \frac{k-\Delta}\tau \tilde y^{Nj} \right) \tilde q^{Nkj} + \frac1{2\pi i \, j^2} \bigl( \tilde y^{Nj} - (\tilde q/ \tilde y)^{Nj} \bigr) \tilde q^{Nkj} \right] \;.
\ee
In the regime of convergence (\ref{regime of convergence}) this series goes to zero as $N \to \infty$. We conclude that the only contribution at leading order in $N$ is from the polynomial $\cQ$ in (\ref{def Q}).

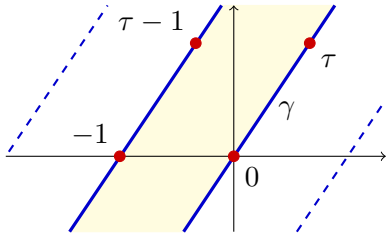
\begin{figure}[t]
\centering
\begin{tikzpicture}
\filldraw [yellow!15!white] (-2.16,-1) to (-.66,-1) to (1.33,2) to (-.16,2) to cycle;
\draw [->] (-3,0) to (2,0);
\draw [->] (0,-1) to (0,2);
\draw [very thick, blue!80!black] (-.66,-1) to (1.33,2);
\draw [very thick, blue!80!black] (-2.16,-1) to (-.16,2);
\draw [dashed, thick, blue!80!black] (-2.96, .06) to (-1.66,2);
\draw [dashed, thick, blue!80!black] (.83, -1) to (2, .75); 
\filldraw [red!80!black] (0,0) circle [radius=.07] node [below right, black] {\small $0$};
\filldraw [red!80!black] (1,1.5) circle [radius=.07] node [below right, black] {\small $\tau$};
\filldraw [red!80!black] (-1.5,0) circle [radius=.07] node [above left, black] {\small $-1$};
\filldraw [red!80!black] (-.5,1.5) circle [radius=.07] node [above left, black] {\small $\tau-1$};
\node at (.7, .6) {\small $\gamma$};
\end{tikzpicture}
\caption{In yellow is highlighted the domain (\ref{regime of convergence}) in the complex $\Delta$-plane. The right boundary is the line $\gamma$, passing through 0 and $\tau$. The left boundary is the line $\gamma - 1$, passing through $-1$ and $\tau - 1$. The dashes lines are other elements of $\gamma + \bZ$.
\label{fig: strip}}
\end{figure}

The limit we computed is valid as long as $\Delta$ satisfies (\ref{regime of convergence}). That inequality has the interpretation that $\Delta$ should lie inside an infinite strip, bounded on the left by the line through $-1$ and $\tau-1$, and on the right by the line (that we dub $\gamma$) through $0$ and $\tau$ (see Figure~\ref{fig: strip}). On the other hand, $\wt\Gamma(u_{ij} + \Delta; \tau, \tau)$ is a periodic function invariant under shifts $\Delta \to \Delta+1$. Therefore, unless $\Delta$ sits exactly on one image of the line $\gamma$ under periodic integer shifts, there always exists a shift that brings $\Delta$ inside the strip. This means that we can use our computation to extract the limit for all $\Delta \in \bC \setminus \{\gamma + \bZ\}$.

Let us define the periodic discontinuous function
\be
[\Delta]_\tau \,\equiv\, \Big( \Delta + n \,\Big|\, n \in \bZ ,\;\; \im\bigl( - \tfrac1\tau \bigr) > \im\bigl( \tfrac{\Delta + n}\tau \bigr) > 0 \Big) \qquad\quad\text{for } \im\bigl( \tfrac\Delta\tau \bigr) \not\in \bZ \times \im\bigl( \tfrac1\tau \bigr) \;.
\ee
The function is not defined for $\im(\Delta/\tau) \in \bZ \times \im(1/\tau)$. Essentially, this function is constructed in such a way that $[\Delta]_\tau = \Delta \mod 1$, and $[\Delta]_\tau$ satisfies (\ref{regime of convergence}) when it is defined. It also satisfies
\be
[\Delta+1]_\tau = [\Delta]_\tau \;,\qquad\qquad [\Delta+\tau]_\tau = [\Delta]_\tau + \tau \;,\qquad\qquad [-\Delta]_\tau = -[\Delta]_\tau - 1 \;.
\ee
We use such a function to express the limit as
\bea
\label{limit single term Gamma tilde}
& \lim_{N\to\infty} \sum_{i\neq j}^N \log \wt\Gamma\bigl( u_{ij} + \Delta; \tau, \tau \bigr) \Big|_{(\ref{BA solution basic})} = - \pi i \sum_{i\neq j}^N \cQ\bigl( u_{ij} + [\Delta]_\tau; \tau, \tau \bigr) + \cO(N) \\
&\qquad = - \pi i N^2 \, \frac{ \big( [\Delta]_\tau - \tau \big)\big( [\Delta]_\tau - \tau + \tfrac12 \big)\big( [\Delta]_\tau - \tau + 1 \big) }{ 3\tau^2} + \cO(N) \;.
\eea
This expression is, by construction, invariant under $\Delta \to \Delta+1$. The lines
\be
\label{Stokes lines in Gamma tilde}
\im( \Delta/\tau) \,\in\, \bZ \times \im(1/\tau)
\ee
that we have dubbed $\gamma+\bZ$, are Stokes lines: they represent transitions between regions in the complex $\Delta$-plane in which different exponential contributions dominate the large $N$ limit, and along which the limit is discontinuous.%
\footnote{Stokes lines divide the complex plane into regions in which the limit gives different analytic functions. Because of their origin, Stokes lines have the property that only the imaginary part of the function can jump, while the real part must be continuous. One can indeed check that (\ref{limit single term Gamma tilde}) satisfies this property.}
We do not know what is the limit along the lines, because different contributions compete and a more precise estimate would be necessary to evaluate their sum. We will elaborate on Stokes lines in Section~\ref{sec: final result}.

The term with $\Delta=0$ requires a special treatment, because it does not satisfy (\ref{regime of convergence}). We can still use the expansion (\ref{modular expansion Gamma tilde}). The term $\log\theta_0$ is evaluated as
\begin{align}
{} & \sum_{i\neq j}^N \log\theta_0 \left( \frac{u_{ij}}\tau; -\frac1\tau \right) = \sum_{i\neq j}^N \sum_{k=0}^\infty \log \left[ \left( 1 - \frac{\tilde z_i}{\tilde z_j} \, \tilde q^k \right) \left( 1 - \frac{\tilde z_j}{\tilde z_i} \, \tilde q^{k+1} \right) \right] \\
&\quad = \sum_{i\neq j}^N \log \left( 1 - \frac{\tilde z_i}{\tilde z_j} \right) + 2 \sum_{i\neq j}^N \sum_{k=1}^\infty \log \left( 1 - \frac{\tilde z_i}{\tilde z_j} \, \tilde q^k \right) 
= N\log N + 2N \log \frac{\bigl( \tilde q^N; \tilde q^N \bigr)_\infty }{ (\tilde q; \tilde q)_\infty} \;. \nn
\end{align}
To calculate the first term in the second line, we notice that $x^N - 1 = \prod_{j=1}^N \big( x - e^{2\pi i j/N} \big)$. Factoring $(x-1)$ on both sides we get $x^{N-1} + \ldots + x + 1 = \prod_{j=1}^{N-1} \big( x - e^{2\pi i j/N} \big)$, and substituting $x=1$ we get $N = \prod_{j=1}^{N-1} \bigl( 1 - e^{2\pi i j/N} \bigr)$. At this point we can shift $j$ by $k$ units and multiply over $k$:
\be
N^N = \prod_{k=1}^N \prod_{j\, (\neq k) \, =1}^N \Big( 1 - e^{2\pi i (j-k)/N} \Big) \;.
\ee
To compute the second term we use the series expansion as before. We see that $\log\theta_0$ contributes at order $\cO(N\log N)$ and can be neglected. The product of terms $\log\psi$ gives
\bea
{} & \sum_{i\neq j}^N \log \prod_{k=0}^\infty \frac{ \psi\big( \frac{k+1 + u_{ij}}\tau \big) }{ \psi\big( \frac{k - u_{ij}}\tau \big)} = \sum_{i\neq j}^N \log \prod_{k=0}^\infty \frac{ \psi\big( \frac{k+1 + u_{ij}}\tau \big) }{ \psi\big( \frac{k + u_{ij}}\tau \big)} = - \sum_{i\neq j}^N \log\psi\Big( \frac{u_{ij}}\tau \Big) \\
&\qquad = \sum_{i<j}^N \pi i \left( \frac{(j-i)^2}{N^2} - \frac16 \right) = \frac{i\pi}{12} (N-1) \;.
\eea
In the first equality we changed sign to $u_{ij}$ because it is summed over $ij$; to go to the second line we used (\ref{properties psi}). This term is of order $\cO(N)$ and can be neglected. We conclude that
\be
\lim_{N\to\infty} \sum_{i\neq j}^N \log \wt\Gamma(u_{ij} ; \tau, \tau) \Big|_{(\ref{BA solution basic})} = \pi i N^2 \, \frac{ \tau \, \big( \tau - \tfrac12 \big)\big( \tau - 1 \big) }{ 3\tau^2} + \cO(N \log N) \;.
\ee

\paragraph{Total contribution from the basic solution.} At this point we can collect the various contributions and obtain the large $N$ limit of $\log\cZ$ in (\ref{Z in BA formula}) evaluated on the solution (\ref{BA solution basic}). The expression depends on $[\Delta_1]_\tau$, $[\Delta_2]_\tau$ and $[\Delta_1 + \Delta_2]_\tau$. We notice the following relation:
\be
\label{def two cases}
[\Delta_1 + \Delta_2]_\tau = \begin{cases} [\Delta_1]_\tau + [\Delta_2]_\tau &\text{if } \im\left( - \frac1\tau\right) > \im \left( \frac{ [\Delta_1]_\tau + [\Delta_2]_\tau}\tau \right) > 0 \hspace{1.85cm} \text{1$^\text{st}$ case} \\
[\Delta_1]_\tau + [\Delta_2]_\tau + 1 &\text{if } \im\left( - \frac2\tau\right) > \im \left( \frac{ [\Delta_1]_\tau + [\Delta_2]_\tau}\tau \right) > \im\left( - \frac1\tau \right) \quad \text{2$^\text{nd}$ case.} \end{cases}
\ee
The second one can be rewritten as
\be
[-\Delta_1 - \Delta_2] _\tau = [-\Delta_1]_\tau + [-\Delta_2]_\tau \qquad \text{if }  \im\left( - \tfrac1\tau\right) > \im \left( \tfrac{ [-\Delta_1]_\tau + [-\Delta_2]_\tau}\tau \right) > 0 \;.
\ee
The large $N$ limit of the summand is then
\be
\label{large N limit Z on basic solution}
\lim_{N\to\infty} \log\cZ \Big|_{(\ref{BA solution basic})} = - \pi i N^2\, \Theta(\Delta_1, \Delta_2; \tau) \;,
\ee
where we have introduced the following function for compactness:
\be
\label{def Theta}
\Theta(\Delta_1, \Delta_2; \tau) = \begin{cases} \dfrac{[\Delta_1]_\tau [\Delta_2]_\tau \big( 2\tau - 1 - [\Delta_1]_\tau - [\Delta_2]_\tau \big)}{\tau^2} &\text{1$^\text{st}$ case} \\[.9em]
\dfrac{\big( [\Delta_1]_\tau +1 \big) \big( [\Delta_2]_\tau +1 \big) \big( 2\tau - 1 - [\Delta_1]_\tau - [\Delta_2]_\tau \big)}{\tau^2} -1 \quad&\text{2$^\text{nd}$ case.} \end{cases}
\ee
The two cases were defined in (\ref{def two cases}).

We can rewrite the function $\Theta$ in a way that will be useful in Section~\ref{sec: entropy from index}. Define an auxiliary chemical potential $\Delta_3$, modulo 1, such that
\be
\label{constraint chemical pot mod 1}
\Delta_1 + \Delta_2 + \Delta_3 - 2 \tau \,\in\, \bZ \;.
\ee
It follows that $[\Delta_3]_\tau = 2\tau - [\Delta_1 + \Delta_2]_\tau - 1$. It is also useful to define the primed bracket
\be
[\Delta]'_\tau = [\Delta]_\tau + 1 \qquad\Rightarrow\qquad 0 > \im\left( \frac{ [\Delta]'_\tau}\tau \right) > \im\left( \frac1\tau\right) \;.
\ee
The primed bracket selects the image of $\Delta$, under integer shifts, that sits inside the strip on the right of the line $\gamma$ through zero and $\tau$, as opposed to the strip on the left. Hence
\be
\label{def Theta alternative}
\Theta(\Delta_1, \Delta_2; \tau) = \begin{cases}
\dfrac{[\Delta_1]_\tau [\Delta_2]_\tau [\Delta_3]_\tau}{\tau^2} \qquad\quad&\text{if} \quad \im\left( - \frac1\tau \right) > \im\left( \frac{[\Delta_1]_\tau + [\Delta_2]_\tau}\tau \right) > 0 \\[1em]
\dfrac{[\Delta_1]'_\tau [\Delta_2]'_\tau [\Delta_3]'_\tau}{\tau^2} - 1 &\text{if} \quad 0 > \im\left( \frac{[\Delta_1]'_\tau + [\Delta_2]'_\tau}\tau \right) > \im\left( \frac1\tau \right)  \;.
\end{cases}
\ee
Irrespective of the integer appearing in (\ref{constraint chemical pot mod 1}), the bracketed potentials satisfy the following constraints:
\bea {}
[\Delta_1]_\tau + [\Delta_2]_\tau + [\Delta_3]_\tau -2\tau + 1 &= 0 \qquad&&\text{1$^\text{st}$ case} \\[.5em]
[\Delta_1]'_\tau + [\Delta_2]'_\tau + [\Delta_3]'_\tau -2\tau - 1 &= 0 \qquad&&\text{2$^\text{nd}$ case} \;.
\eea
Such constraints have already appeared in \cite{Hosseini:2017mds, Cabo-Bizet:2018ehj, Choi:2018hmj}.

\subsection[Contribution of $SL(2,\bZ)$-transformed solutions]{Contribution of \matht{SL(2,\bZ)}-transformed solutions}

As discussed in Section~\ref{sec: exact solutions}, (\ref{BA solution basic}) is not the only solution to the BAEs: each inequivalent $SL(2,\bZ)$ transformation of it, given in (\ref{BA solutions SL transformed}), is another solution---and even more generally there are the $\{m,n,r\}$ solutions (\ref{BA solutions Hong-Liu}) found in \cite{Hong:2018viz}. Some of those solutions might contribute at the same leading order in $N$

A class of inequivalent solutions---particularly simple to study---that contribute at leading order in $N$ is obtained through $T$-transformations:
\be
\label{BA solution T-transformed}
u_{ij} = \frac{\tau + r}N \, (j-i) \qquad\qquad\text{for } r = 0, \dots, N-1 \;.
\ee
These are the solutions $\{1,N,r\}$ in the notation of Section~\ref{sec: exact solutions}. To evaluate their contribution, simply notice that both $\cZ$ in (\ref{Z in BA formula}) and $H$ are invariant under $\tau \to \tau+r$, thus the contribution of (\ref{BA solution T-transformed}) is the same as in (\ref{large N limit Z on basic solution}) but with $\tau \to \tau + r$. In the large $N$ limit, $r$ runs over $\bZ$.

We have not evaluated the contribution of all other $\{m,n,r\}$ solutions, which is a difficult task. However, in order to have an idea of what their contribution could be, let us estimate the contribution from the $S$-transformed solution
\be
u_{ij} = \frac{j-i}N \;,
\ee
which is $\{N,1,0\}$ in the notation of (\ref{BA solutions Hong-Liu}). The large $N$ limit of $\kappa_N$ does not depend on the solution, and is subleading. The large $N$ limit of $\log H$  is computed in the same way as in Section~\ref{sec: contrib basic sol}, and it gives $\cO(N\log N)$ or smaller. Let us then analyze $\cZ$. In the regime $|q|^2 < |y|<1$ we can directly expand $\log\wt\Gamma$ in its plethystic form:
\bea
{} & \sum_{i\neq j}^N \log \wt{\Gamma}(u_{ij} + \Delta; \tau,\tau) = \sum_{i\neq j}^N \sum_{\ell=1}^\infty \sum_{m=0}^\infty \frac{m+1}\ell \left( \left(\frac{z_i}{z_j}\right)^\ell y^\ell - \left( \frac{z_j}{z_i} \right)^\ell y^{-\ell} q^{2\ell} \right) q^{m\ell} \\
&\quad = \sum_{\ell=1}^\infty \sum_{m=0}^\infty \frac{m+1}\ell \, A_\ell \, \bigl( y^\ell - y^{-\ell} q^{2\ell} \bigr) \, q^{m\ell} = N \log \frac{ \wt\Gamma(N\Delta; N\tau, N\tau) }{ \wt\Gamma(\Delta; \tau, \tau) } = \cO(N) \;.
\eea
If $|y|$ is outside the range of convergence of the plethystic expansion, either above or below, we can simply shift $\Delta \to \Delta \pm \tau$. This gives a shift by
\be
\pm \sum_{i\neq j}^N \log\theta_0(u_{ij} + \Delta; \tau) = \cO(N)
\ee
which can be treated in a similar way. This confirms that the estimate above is valid for all $\Delta$'s, even outside the original regime of convergence. The case $\Delta=0$ requires a special treatment. We have
\bea
{} & \sum_{i\neq j}^N \log \wt{\Gamma}(u_{ij}; \tau, \tau) = \sum_{i\neq j}^N \left[ - \log\left( 1 - \frac{z_i}{z_j} \right) + 2 \sum_{\ell,m=1}^\infty \frac1\ell \left( \frac{z_i}{z_j}\right)^\ell q^{m\ell} \right] \\
&\quad = - N\log N + 2 \sum_{\ell, m=1}^\infty \frac{A_\ell}\ell \, q^{m\ell} = -N\log N + 2N \log \frac{(q;q)_\infty}{(q^N; q^N)_\infty} = \cO(N\log N) \;.
\eea
Thus, there is no contribution from $\log \cZ$ at leading order in $N$.

In the following we will assume that the only solutions contributing at leading order, namely $\cO(N^2)$, are the $T$-transformed solutions.

\subsection{Final result and Stokes lines}
\label{sec: final result}

Since we end up with competing exponentials, the one with the largest real part dominates the large $N$ limit. Assuming that solutions other than the $T$-transformed of the basic one are of subleading order in $N$, we find the final formula
\be
\label{index large N final}
\lim_{N\to \infty} \log \, \cI(q, y_1, y_2) = \wt{\max\limits_{r \,\in\, \bZ}} \, \Bigl( - \pi i N^2 \, \Theta(\Delta_1, \Delta_2; \tau+ r) \Bigr) \;\equiv\; \log\,\cI_\infty \;.
\ee
The function $\Theta$ is defined in (\ref{def Theta}).
The meaning of $\wt\max$ is that we should choose the value of $r \in \bZ$ such that the real part of the argument is maximized. One of the good features of eqn.~(\ref{index large N final}) is that it is periodic under integer shifts of $\tau, \Delta_1, \Delta_2$. We already observed that $\Theta$ is periodic in $\Delta_{1,2}$ because the functions $[\Delta_{1,2}]_\tau$ are. Taking the $\wt\max$ over $\tau \to \tau + r$ gives periodicity in $\tau$ as well. This implies that the RHS of (\ref{index large N final}) is actually a single-valued function of the fugacities $q,y_1, y_2$. This is a property of the index at finite $N$, as manifest in (\ref{eq:traceindex y}) and (\ref{eq:BAindex}), and it is reassuring that the large $N$ expression we found respects the same property.

The function $\cI_\infty$ has a complicated structure. The full range of allowed fugacities $q,y_1, y_2$ gets divided into multiple domains of analyticity, separated by \emph{Stokes lines}. In each domain of analyticity, only one exponential contribution (for some value of $r$) dominates the large $N$ limit: the function $\log\cI_\infty$ takes the form of a simple rational function given by $\Theta(\Delta_1, \Delta_2; \tau+r)$. The Stokes lines are real-codimension-one surfaces, in the space of fugacities, that separate the different domains. When crossing a Stokes line, a different exponential contribution dominates, and $\log\cI_\infty$ takes the form of a different rational function. In particular, on top of a Stokes line there are two (or more) exponential contributions that compete: their exponents have equal real part. This characterizes the locations of Stokes lines. In terms of the function $\Theta$:
\be
\im \Theta(\Delta_1, \Delta_2; \tau+r_1) = \im \Theta(\Delta_1, \Delta_2; \tau + r_2)
\ee
for some $r_{1,2} \in \bZ$.

In fact, also the values of $\Delta_1, \Delta_2$ such that $\Theta(\Delta_1, \Delta_2; \tau+r)$ is discontinuous (for the value of $r$ picked up by $\wt\max$) should be regarded as forming a Stokes line. In this case, the two competing exponents correspond to the values of $\Theta$ on the two sides of the discontinuity. There are two possible sources of discontinuity. First, one of the bracket functions, say $[\Delta_1]_\tau$, could be discontinuous. This happens when $\im(\Delta_1/\tau) \in \bZ \times \im(1/\tau)$, namely when $\alpha \,\equiv\, \lim_{\epsilon \to 0^+} [\Delta_1-\epsilon]_\tau/\tau \in \bR$. Taking into account that on the left of the discontinuity we are in the 1$^\text{st}$ case, while on the right we are in the 2$^\text{nd}$ case---in the terminology of (\ref{def two cases})---and assuming that $\Delta_2$ is generic, we find
\be
\lim_{\epsilon\to 0^+} \biggl[ \Theta(\Delta_1 - \epsilon, \Delta_2; \tau) - \Theta(\Delta_1 + \epsilon, \Delta_2;\tau) \biggr] = (\alpha-1)^2 \;\in\; \bR \;,
\ee
where the limit is taken with $\epsilon$ real positive. Second, we could pass from the 1$^\text{st}$ to the 2$^\text{nd}$ case of the definition (\ref{def Theta}). This happens when $[\Delta_1]_\tau + [\Delta_2]_\tau + 1 = \alpha \, \tau$ for some $\alpha\in \bR$. Assuming that $\Delta_{1,2}$ are otherwise generic, we find
\be
\Delta\Theta = (\alpha-1)^2 \;\in\; \bR \;.
\ee
In both cases we confirm that the codimension-one surface of discontinuity is a Stokes line, because $\im\Theta$ is equal on the two sides.

When we sit exactly on a Stokes line, two (or more) exponential contributions compete, and in order to compute the large $N$ limit we should sum them. However we do not know the relative phases, because they are affected by all subleading terms and a more accurate analysis would be required. Therefore, we cannot determine the large $N$ limit of the index along Stokes lines.

It turns out that a value of $r$ that maximizes the real part of the argument of $\wt\max$ may or may not exist. We can estimate the behaviour of the real part at large $r$ by noticing that
\be
\lim_{r\to \pm\infty} \frac{ [\Delta]_{\tau+r}}{\tau+r} = \frac{\im\Delta}{\im\tau} \;.
\ee
This implies that
\be
\label{limiting value im Theta}
\lim_{r\to\pm\infty} \im \Theta(\Delta_1, \Delta_2; \tau+r)= \frac{\im\Delta_1\, \im\Delta_2\, \im(2\tau-\Delta_1 - \Delta_2)}{(\im\tau)^2} \;.
\ee
Thus, the real part of the argument of $\wt\max$ approaches a constant value. If there is no maximum but rather the constant value is a supremum, then our computation is not finished:  All contributions from the $T$-transformed solutions should be summed, however for large $|r|$ they form an infinite number of competing exponentials, whose sum crucially depends on how they interfere. In order to determine such a sum we would need more accurate information.

We conclude by stressing that---even though only the dominant exponential determines the large $N$ limit of the index---we expect that all exponential contributions, including the subdominant ones, have some physical meaning. Each of them plays the role of a ``saddle point'', although our treatment is not the standard saddle-point approximation. We will make this comment more concrete in Section~\ref{sec: entropy from index}, when comparing the large $N$ limit of the index with BPS black hole solutions in supergravity.

\subsection{Comparison with previous literature}

The large $N$ limit of the superconformal index of $\cN=4$ SYM was already computed in \cite{Kinney:2005ej}. There, it was found that the large $N$ limit does not depend on $N$, and therefore it does not show a rapid enough growth of the number of states to reproduce the black hole entropy. In this Section we would like to explain how the results here and there can be compatible.

The authors of \cite{Kinney:2005ej} took the large $N$ limit of the index, for \emph{real} fugacities. Their result, in our notation and restricted to the case $p=q$, is
\be
\label{result of Maldacena}
\lim_{N \to \infty} \cI(q,y_1,y_2) = \prod_{n=1}^\infty \frac1{1-f\bigl( q^n, y_1^n, y_2^n\bigr)}
\ee
with
\be
1-f(q,y_1,y_2) = \frac{(1-y_1)(1-y_2)(1-q^2/y_1y_2)}{(1-q)^2} \;.
\ee
In particular, $\log\cI$ is of order $\cO(1)$. On the contrary, we computed the large $N$ limit for generic \emph{complex} fugacities, and found that $\log\cI$ is of order $\cO(N^2)$. It was already discussed in \cite{Choi:2018hmj}, in a double-scaling Cardy-like limit, that the large $N$ limit of the index is completely different for real and complex fugacities, and it was observed in \cite{Choi:2018vbz} that there exists a deconfinement transition once complex fugacities are taken into account.

The resolution we propose relies on the fact that, for complex fugacities, the limit shows Stokes lines. As we described, along those codimension-one surfaces multiple exponentials compete. In order to know what the limit is there, we would need to sum those competing exponentials, but this requires a more accurate knowledge of the subleading terms.

What we notice, though, is that the codimension-three subspace of real fugacities is precisely within a Stokes line. Therefore, although we cannot prove it, it is conceivable that the competing terms cancel exactly, leaving the $\cO(1)$ result (\ref{result of Maldacena}). Indeed, in Appendix~\ref{app: real fugacities} we prove the following result, which is stronger than the statement that we sit on a Stokes line. Take the angular fugacity $q$ to be real positive, namely $0<q<1$ and set $\tau \in i\bR_{\geq0}$ for concreteness, and take the flavor fugacities $y_{1,2}$ to be real. Then $\Theta(\Delta_1, \Delta_2; \tau)$ is along a Stokes line and is not defined, while
\be
\Theta(\Delta_1, \Delta_2; \tau-r) = - \, \wb{\Theta(\Delta_1, \Delta_2; \tau+r)} \, - 1
\ee
for $r>0$. On the other hand, take the angular fugacity real negative, namely $-1<q<0$ and set $\tau \in - \frac12 + i\bR_{\geq0}$, and take again the flavor fugacities to be real. Then
\be
\Theta(\Delta_1, \Delta_2; \tau-r) = - \, \wb{\Theta(\Delta_1, \Delta_2; \tau+r+1)} \, - 1
\ee
for $r\geq 0$. Therefore, among the various contributions from $T$-transformed solutions parameterized by $r\in\bZ$, there is an exact pairing of all well-defined terms where, in each pair, two terms have the same real part and can conceivably cancel. In other words, not only the term with maximal real part can cancel, but also all other terms we computed at order $\cO(N^2)$. This scenario is a strong check of our result, that makes it compatible with \cite{Kinney:2005ej}.

\section{Statistical interpretation and \matht{\cI}-extremization}
\label{sec: stat interpretation}


We wish to extract the number of BPS states, for given electric charges and angular momenta, from the large $N$ limit of the exact expression (\ref{eq:BAindex}) of the superconformal index. Since the latter counts states weighted by the fermion number $(-1)^F$, one may worry that strong cancelations take place and that the total number of states is not accessible. However, one can argue \cite{Benini:2015eyy, Benini:2016rke} that the index (\ref{eq:traceindex}) or (\ref{eq:traceindex y}) is equal to
\be
\label{index with R_trial}
\cI(p,q,y_1, y_2) = \Tr \, e^{i\pi R_\text{trial}(\tau,\sigma,\Delta_1,\Delta_2)} \; e^{-2\pi \im \left[ \rule{0pt}{.7em} \tau C_1 + \sigma C_2 + \Delta_1 C_3 + \Delta_2 C_4 \right] } \; e^{-\beta\{\cQ, \cQ^\dag\}} \;,
\ee
where the trace is taken in the IR $\cN=2$ super quantum mechanics (QM) obtained by reducing the 4d theory on $S^3$, $R_\text{trial}$ is a trial R-symmetry, and $C_{1,2,3,4}$ are the charges appearing in (\ref{eq:traceindex y}):
\be
C_1 = J_1 + \frac{R_3}2 \;,\qquad C_2 = J_2 + \frac{R_3}2 \;,\qquad C_3 = q_1 \;,\qquad C_4 = q_2 \;.
\ee
Indeed, because of the relations (\ref{relation spin F R}), we can represent the fermion number as $(-1)^F= e^{i \pi R_3}$. Substituting in (\ref{eq:traceindex y}) and separating the chemical potentials into real and imaginary part, we obtain the expression (\ref{index with R_trial}) with
\be
R_\text{trial}(\tau,\sigma,\Delta_1,\Delta_2) = R_3 + 2\re\tau \, C_1 + 2 \re\sigma \, C_2 + 2\re\Delta_1 \, C_3 + 2\re\Delta_2 \ C_4 \;.
\ee
From the point of view of the super QM, $R_3$ is an R-symmetry while the other four operators are flavor charges, hence $R_\text{trial}$ is an R-symmetry. We see in (\ref{index with R_trial}) that only the first exponential can produce possibly-dangerous phases, while the other two are real positive.

Now, for a single-center black hole in the microcanonical ensemble, the near-horizon AdS$_2$ region is dual to an $\cN=2$ superconformal QM. The black hole states are vacua of the $\fsu(1,1|1)$ 1d SCA. Since we are in the microcanonical ensemble, each of those states is invariant under the global conformal algebra $\fsu(1,1) \cong \fso(2,1)$ (because AdS$_2$ is) as well as under the fermionic generators (because the black hole is supersymmetric). This necessarily implies that those states are invariant under the superconformal R-symmetry $\fu(1)_\text{sc} \subset \fsu(1,1|1)$, \ie{} that they have vanishing IR superconformal R-charge $R_\text{sc}$. Thus, when $R_\text{trial}$ is tuned to $R_\text{sc}$, the index counts the black hole states with no extra signs or phases (this is similar to \cite{Sen:2009vz, Dabholkar:2010rm}). Of course, in a given charge sector there will be more BPS states than just the single-center black hole, but assuming that the single-center black hole dominates, the index captures its entropy. It remains to understand how to identify $R_\text{sc}$. At large $N$, the entropy is extracted from the index with a Legendre transform, and this operation can be argued to effectively select $R_\text{sc}$ among the $R_\text{trial}$'s (this large $N$ principle was dubbed $\cI$-extremization in \cite{Benini:2015eyy, Benini:2016rke}).

Let us elaborate on this point. The index is the grand canonical partition function of BPS states. Introducing an auxiliary variable $\Delta_3$ and the corresponding fugacity $y_3 = e^{2\pi i \Delta_3}$ such that $\Delta_1 + \Delta_2 + \Delta_3 - \tau - \sigma +1 \in 2\bZ$, we can rewrite (\ref{index with R_trial}) as
\be
\label{index full chem pot}
\cI(p,q,y_1,y_2) = \Tr_\text{BPS} \, p^{J_1} \, q^{J_2} \, y_1^{Q_1} \, y_2^{Q_2} \, y_3^{Q_3} \;.
\ee
Here the trace is over states with $\{\cQ, \cQ^\dag\}=0$, and we have identified $Q_I = R_I/2$ (for $I=1,2,3$) with the electric charges in supergravity. We recognise that the black hole angular momenta $J_{1,2}$ are associated with the chemical potentials $\tau,\sigma$ and the charges $Q_{1,2,3}$ with $\Delta_{1,2,3}$. The microcanonical degeneracies at fixed quantum numbers are extracted by computing the Fourier transform of (\ref{index full chem pot}). However, since $\Delta_3$ is not an independent variable, what we obtain are the degeneracies for fixed values of the four charge operators appearing in (\ref{eq:traceindex y}), summed over $Q_3$. Using the supergravity notation, those four fixed charge operators are
\be
C_1 = J_1 + Q_3 \;,\qquad C_2 = J_2 + Q_3 \;,\qquad C_3 = Q_1 - Q_3 \;,\qquad C_4 = Q_2 - Q_3 \;.
\ee
Thus, what we can compute is
\be
\label{eq:dmicro}
\sum_{Q_3} d(J,Q) \Big|_{C_{1,2,3,4}} = \int \! d\tau \, d\sigma \, d\Delta_1 \, d\Delta_2 \;\; \cI(p,q,y_1,y_2) \; p^{-J_1}q^{-J_2}\prod_{I=1}^3 y_I^{-Q_I}
\ee
where $d(J,Q)$ are the (weighted) degeneracies with all charges $J_{1,2}$ and $Q_{1,2,3}$ fixed.

Nevertheless, we can take advantage of the fact, reviewed in Section~\ref{sec: BHs}, that the charges of BPS back holes are constrained, and for fixed $C_{1,2,3,4}$ there is at most one black hole---for a certain value of the fifth charge $Q_3$. We can then use (\ref{eq:dmicro}) to extract its degeneracy $d(J,Q) = \exp S_\text{BH}(J,Q)$ at leading order because the latter will dominate the sum over $Q_3$.

In the large $N$ limit, the integral \eqref{eq:dmicro} reduces by saddle point approximation to a Legendre transform with respect to the independent variables $\{\tau, \sigma, \Delta_1, \Delta_2\}$:
\bea
\label{eq:Smicro}
S_\text{BH}(J,Q) &= \log \cI\bigl( \wh\tau,\wh\sigma, \wh\Delta_1, \wh\Delta_2 \bigr) - 2\pi i\Bigl(\wh\tau J_1 + \wh\sigma J_2 +\sum_{I=1}^3\wh\Delta_I Q_I \Bigr) \\
&= \log \cI\bigl( \wh\tau,\wh\sigma, \wh\Delta_1, \wh\Delta_2 \bigr) - 2\pi i\Bigl(\wh\tau C_1 + \wh\sigma C_2 + \wh\Delta_1 C_3 + \wh\Delta_2 C_4 \Bigr) + 2\pi i Q_3 \;,
\eea
where hatted variables denote the critical point. In this approach, $Q_3$ can be determined as the unique value that makes the entropy $S_\text{BH}(J,Q)$ real \cite{Benini:2015eyy, Benini:2016rke}.

In the particular case of 4d $\cN=4$ SYM, the large $N$ limit of the index is a function with multiple domains of analyticity, separated by Stokes lines. This makes things more interesting. In each domain we should perform the Legendre transform, and whenever the critical point falls inside the domain itself, we obtain a self-consistent contribution to the total entropy. Even more generally, we have written the index as a sum of competing exponentials (one for each Bethe Ansatz solution) and we can compute the Legendre transform of each of those exponentials---irrespective of which one dominates. We expect each contribution to represent the entropy of some classical solution---very similarly to a standard saddle point---even when the entropy is smaller than that of the dominant solution.

\section{Black hole entropy from the index}
\label{sec: entropy from index}

In this Section we show that the contribution of the basic solution (\ref{BA solution basic}) to the superconformal index at large $N$, in the domain of analyticity that we called ``1$^\text{st}$ case'' in (\ref{def two cases}), given by
\be
\label{sector of BH index}
- \pi i N^2 \, \Theta(\Delta_1, \Delta_2; \tau) \Big|_\text{1$^\text{st}$ case} = - \pi i N^2 \, \frac{[\Delta_1]_\tau [\Delta_2]_\tau \bigl( 2\tau - 1 - [\Delta_1]_\tau - [\Delta_2]_\tau \bigr) }{ \tau^2 }
\ee
precisely reproduces the Bekenstein-Hawking entropy (\ref{eq:BHentropy J1=J2}) of single-center black holes in AdS$_5$ (this is in line with the result of \cite{Choi:2018hmj} in a double-scaling Cardy-like limit). It amounts to show that the Legendre transform of (\ref{sector of BH index}) is the black hole entropy (this will be reviewed below), and that the critical point involved in the Legendre transform consistently lies within the domain of analyticity in which (\ref{sector of BH index}) holds.

Recall that the contribution of the basic solution corresponds to the $r=0$ sector in (\ref{index large N final}). For black holes with large charges, \ie{} for black holes that are large compared with the AdS$_5$ scale, that is indeed the dominant contribution to the index. However, intriguingly enough, as we reduce the charges the contribution of the single-center black hole may cease to dominate. We will highlight this phenomenon in Section~\ref{sec: case study} in the very special case of black holes with equal charges. This seems to suggest that, below a certain threshold, the BPS black holes may develop instabilities, possibly towards hairy or multi-center black holes. Indications that this is the case have also been given in \cite{Choi:2018hmj, Choi:2018vbz}. It would be nice if there was a connection between this observation and recently constructed hairy black holes in AdS$_5$ \cite{Bhattacharyya:2010yg, Markeviciute:2016ivy, Markeviciute:2018yal, Markeviciute:2018cqs}.

\paragraph{The entropy function.}
The Legendre transform of the black hole entropy (\ref{eq:BHentropy}) in the general case, also called entropy function, was obtained in \cite{Hosseini:2017mds}. Let us review it, following the detailed discussion in Appendix~B of \cite{Cabo-Bizet:2018ehj}. The entropy function is
\be
\label{entropy function Zaffaroni}
\cS = -2\pi i \nu \, \frac{X_1 \,  X_2 \, X_3}{\omega_1 \, \omega_2} \qquad\qquad\text{with}\qquad \nu = \frac{N^2}2 = \frac{\pi}{4G_N g^3}
\ee
and with the constraint
\be
\label{constraint on chem pot Zaffaroni}
\sum_{a=1,2,3} X_a - \sum_{i=1,2} \omega_i + 1 = 0 \;.
\ee
Because of the constraint, $\cS$ is really a function of four variables. The entropy $S_\text{BH}$ is the Legendre transform of $\cS$ with its constraint. We can compute it as the critical point of
\be
\label{entropy before extremization}
\wh\cS = \cS - 2\pi i \left( \sum\nolimits_a Q_a \, X_a + \sum\nolimits_i J_i \, \omega_i \right) - 2\pi i \Lambda \left( \sum\nolimits_a X_a - \sum\nolimits_i \omega_i + 1 \right)
\ee
in which the constraint is imposed with a Lagrange multiplier $\Lambda$. The equations for the critical point are
\be
Q_a + \Lambda = \frac1{2\pi i} \, \parfrac{\cS}{X_a} \;,\qquad\qquad J_i - \Lambda = \frac1{2\pi i} \, \parfrac{\cS}{\omega_i} \;,
\ee
and the constraint (\ref{constraint on chem pot Zaffaroni}). In details,
\bea
\label{eqns from extremization}
Q_1 + \Lambda &= - \nu \, \frac{X_2 \, X_3}{\omega_1 \, \omega_2} \;,\qquad& Q_2 + \Lambda &= - \nu \, \frac{X_1 \, X_3}{\omega_1 \, \omega_2} \;,\qquad& Q_3 + \Lambda &= - \nu \, \frac{X_1 \, X_2}{\omega_1 \, \omega_2} \\[.5em]
J_1 - \Lambda &= \nu \, \frac{X_1 \, X_2 \, X_3}{\omega_1^2 \, \omega_2} \;,\qquad& J_2 - \Lambda &= \nu \, \frac{X_1 \, X_2 \, X_3}{\omega_1 \, \omega_2^2} \;.
\eea
It follows that
\be
\label{cubic equation for Lambda}
0 = (Q_1+\Lambda) (Q_2 + \Lambda)(Q_3 + \Lambda) + \nu(J_1 - \Lambda) (J_2 - \Lambda) = \Lambda^3 + p_2 \Lambda^2 + p_1\Lambda + p_0
\ee
with
\bea
p_2 &= Q_1 + Q_2 + Q_3 + \nu \\
p_1 &= Q_1Q_2 + Q_1Q_3 + Q_2Q_3 - \nu(J_1 + J_2) \\
p_0 &= Q_1 Q_2 Q_3 + \nu J_1 J_2 \;.
\eea
It turns out that we can find the value of $\wh\cS$ at the critical point without knowing the exact solution for the critical point. We use the fact that $\cS$ is homogeneous of degree 1 (it is a monomial), and thus
\be
\sum\nolimits_a X_a \, \parfrac{\cS}{X_a} + \sum\nolimits_i \omega_i \, \parfrac{\cS}{\omega_i} = \cS \;.
\ee
Substituting into (\ref{entropy before extremization}) we find
\be
S_\text{BH} = \wh\cS \Big|_\text{crit} = - 2\pi i \Lambda \;.
\ee
Since $\Lambda$ is the solution to the cubic equation (\ref{cubic equation for Lambda}), it looks like there are three possible values for the entropy. However, since for real charges the cubic equation has real coefficients, we either find 3 real roots or 1 real and 2 complex conjugate roots for $\Lambda$. Imposing that the entropy be \emph{real positive}, we require that there is 1 real and 2 imaginary conjugate roots, then only one of them---the one along the positive imaginary axis---leads to an acceptable value for the entropy. Since $(\Lambda - \beta)(\Lambda - i\alpha)(\Lambda + i\alpha) = \Lambda^3 - \beta \Lambda^2 + \alpha^2\Lambda - \beta\alpha^2$, we obtain the following constraint on the charges:
\be
\label{constraint on p_012}
p_0 = p_1 p_2 \qquad\text{ and }\qquad p_1>0 \;.
\ee
One can check that the parameterization (\ref{eq:charges}) automatically solves the first equation.
Then the roots of (\ref{cubic equation for Lambda}) are $\Lambda \in \{-p_2, \pm i\sqrt{p_1}\}$. The physical solution is
\be
\Lambda = i \sqrt{p_1} \qquad\Rightarrow\qquad S_\text{BH} = 2\pi \sqrt{p_1} \;,
\ee
which is precisely eqn.~(\ref{eq:BHentropy}). We stress that the conditions (\ref{constraint on p_012}) are necessary, but not sufficient, to guarantee that the supergravity solution is well-defined.%
\footnote{As an example, take $Q_1 = Q_2 = Q_3 \equiv Q$ and $J_1 = J_2 \equiv J$. The first equation in (\ref{constraint on p_012}) is solved by $J = -3Q -1 \pm (2Q+1)^{3/2}$, and both branches are covered by the parameterization (\ref{eq:charges}) as $Q = \mu + \mu^2/2$, $J = 3\mu^2/2 + \mu^3$. Then $p_1 = 3Q^2+6Q+2 \mp 2(2Q+1)^{3/2}$ and one can check that, for $Q>0$, both branches have $p_1>0$. However, only the branch with upper sign satisfies also (\ref{eq:BHconstraint 2})---here $\mu>0$---and corresponds to well-defined supergravity solutions, while the branch with lower sign does not.}

It is not difficult to write the values of the chemical potentials at the critical point. To simplify the notation, let us define
\be
P_{1,2,3} = Q_{1,2,3} + \Lambda \;,\qquad P_{4,5} = J_{1,2} - \Lambda \;,\qquad \Phi_{1,2,3} = X_{1,2,3} \;,\qquad \Phi_{4,5} = - \omega_{1,2}
\ee
and use an index $A=1,\dots, 5$. The equations (\ref{eqns from extremization}) imply that
\be
\Phi_A P_A \quad\text{are all equal for $A=1,\dots, 5$} \;.
\ee
Implementing the constraint (\ref{constraint on chem pot Zaffaroni}), the solution is
\be
\label{critical point Legendre}
\Phi_A = - \frac1{P_A} \left( \sum_{B=1}^5 \frac1{P_B} \right)^{-1} \;.
\ee
Since, even for real charges, the $P_A$'s are complex, the solutions $\Phi_A$ are in general complex.

\paragraph{Equal angular momenta.} Let us specialize the formulas to the case $J_1 = J_2 \equiv J$, and determine useful inequalities satisfied by the chemical potentials at the critical point. First of all, from the constraint (\ref{constraint on chem pot Zaffaroni}) it immediately follows
\be
-\frac1\omega = \frac{X_1}\omega + \frac{X_2}\omega + \frac{X_3}\omega - 2 \;.
\ee
At the critical point (\ref{critical point Legendre}) one finds
\be
\frac{X_a}{\omega} = - \frac{J-\Lambda}{Q_a + \Lambda} \;,\qquad\qquad \im \left( \frac{X_a}{\omega} \right) = \sqrt{p_1} \; \frac{Q_a + J}{Q_a^2 + p_1} > 0 \;.
\ee
To obtain the last inequality we used that $Q_a + J>0$ for the BPS black holes, as we showed in (\ref{bounds on BH charges 2}). This implies that
\be
\label{constraint X_a in strip}
\im\left( - \frac1\omega \right) > \im\left( \frac{X_a}\omega \right) > 0 \qquad\qquad\text{for } a = 1,2,3 \;.
\ee
Using the explicit parameterization (\ref{charges for J1=J2}) presented in Section~\ref{sec: BHs} (and setting $g=1$ for the sake of clarity), one can also show that
\be
\re(\omega) = \frac1{2\,(1+\gamma_1)} \;,\qquad\qquad \im(\omega) = \frac{\nu \, \gamma_2}{4\,(1+\gamma_1) \, \sqrt{p_1}} \;,
\ee
where $p_1 = \nu^2\left( (1+\gamma_1)\gamma_3 - \frac14 \gamma_2^2 \right)$.
In particular, the first equation shows that
\be
0 < \re(\omega) < \frac12 \;.
\ee

\paragraph{Entropy from the index.} Finally, we compare the contribution to the index from the basic solution in the 1$^\text{st}$ case, given in (\ref{sector of BH index}), with the entropy function $\cS$ in (\ref{entropy function Zaffaroni}). The latter, after eliminating $X_3$ with the constraint (\ref{constraint on chem pot Zaffaroni}) and restricting to equal angular fugacities, reads
\be
\cS = - \pi i N^2 \, \frac{X_1 \, X_2 \, \bigl( 2\omega - 1 - X_1 - X_2 \bigr) }{ \omega^2} \;.
\ee
We see that it is exactly equal to (\ref{sector of BH index}), as long as we can identify
\be
\tau = \omega \;,\qquad\qquad [\Delta_a]_\tau = X_a \qquad\text{ for }\quad a = 1,2,3 \;.
\ee
This is not obvious, but we can check that it is indeed possible. First of all, $X_1$ and $X_2$ should satisfy the strip inequalities that $[ \,\cdot\,]_\tau$ does, at least in a neighbourhood of the critical point. This is precisely what we proved in (\ref{constraint X_a in strip}). Second, the fugacities at the critical point should also satisfy the inequalities (\ref{def two cases}) that define the 1$^\text{st}$ case. Because of the constraint, this is the same as requiring that also $X_3$ satisfies (\ref{constraint X_a in strip}), which is true. Thus, this concludes our proof. Let us stress that, in our approach, the constraint (\ref{constraint on chem pot Zaffaroni}) with the correct constant term simply comes out of the large $N$ limit.

\bigbreak

One could wonder what is the physics described by the domain of analyticity named $2^\text{nd}$ case in (\ref{def Theta}). It appears that it reproduces the very same black hole entropy as the 1$^\text{st}$ case. Indeed, as apparent from (\ref{def Theta alternative}), in the two cases $\Theta$ takes almost the same form, the only difference being that $[\,\cdot\,]_\tau$ and $[\,\cdot\,]'_\tau$ satisfy opposite strip inequalities and a constraint with opposite constant term. It was already observed in \cite{Cabo-Bizet:2018ehj} that the entropy function $\cS$ reproduces the black hole entropy with either one of the two constraints imposed. We leave for future work to understand what is the role of such a twin contribution.

\subsection{Example: equal charges and angular momenta}
\label{sec: case study}

In order to make some of the previous statements more concrete, we now study in detail a very special case in which the index counts states with equal charges $Q_1 = Q_2 = Q_3 \equiv Q$ and angular momenta $J_1 = J_2 \equiv J$. This will be instructive to elucidate the structure of Stokes lines.

Let us first quickly summarize the properties of black holes and their entropy in this case \cite{Gutowski:2004ez}. We set $\nu=1$ (all charges are in ``units'' of $\nu$) so that
\be
p_0 = Q^3 + J^2 \;,\qquad p_1 = 3Q^2 - 2J \;,\qquad p_2 = 3Q+1 \;,
\ee
and the charge constraint is
\be
p_1 p_2 - p_0 = 8Q^3 + 3Q^2 - 2 (3Q+1)J - J^2 = 0 \;.
\ee
This is quadratic in $J$ and potentially leads to two branches of solutions. However only one of them satisfies (\ref{eq:BHconstraint 2}) when parameterized in terms of $\mu$ (we also set $g=1$):
\bea
Q &= \mu + \frac12 \mu^2 \hspace{5cm} \Lambda = i\sqrt{p_1} \;,\qquad S = 2\pi \sqrt{p_1} \\
J &= (2Q+1)^{3/2} - 3Q -1 = \frac32 \mu^2 + \mu^3 \\
p_1 &= 3Q^2 + 6Q + 2 - 2(2Q+1)^{3/2} = \mu^3 + \frac34 \mu^4 \;.
\eea
The entropy is positive for $Q>0$, and in this range $J>0$.

The extremization problem \eqref{entropy before extremization} simplifies because we only have two chemical potentials, $X\equiv X_1 = X_2 = X_3$ and $\omega$ with the constraint \eqref{constraint on chem pot Zaffaroni}. The critical point is
\be
\label{critical points Q J +1}
\omega = \frac{Q+\Lambda}{2Q + 3J -\Lambda}\;,\qquad\qquad X = - \frac{J-\Lambda}{2Q + 3J +\Lambda} \;.
\ee
Let us mention that in the alternative extremization problem in which the constraint (\ref{constraint on chem pot Zaffaroni}) is modified by changing $+1$ into $-1$, the critical values of $X$ and $\omega$ are given by the same expressions, however the critical value of the Lagrange multiplier becomes $\Lambda=-i\sqrt{p_1}$.

We now turn to the index. Given the identifications $X_a = [\Delta_a]_\tau$ and $\omega = \tau$, we can restrict to chemical potentials such that $[\Delta_1]_\tau=[\Delta_2]_\tau=[\Delta_3]_\tau\equiv [\Delta]_\tau$, where $\Delta_3$ is defined through the general constraint (\ref{constraint chemical pot mod 1}). Up to integer shifts, this amounts to
\be
\Delta_1 = \Delta_2 = \Delta_3 \,\equiv\, \Delta = \frac{2\tau - 1}3 \;.
\ee
The critical points (\ref{critical points Q J +1}) indeed satisfy this relation. We have thus reduced to a single independent chemical potential $\tau$. Notice that the function $\cI\bigl( \Delta(\tau);\tau \bigl) = \cI\bigl( \frac{2\tau -1}3; \tau \bigr)$ is periodic under $\tau \to \tau +3$, therefore we will restrict to $0 \leq \re\tau<3$.

We study the large $N$ formula \eqref{index large N final} for the index, in particular we want to determine the structure of the leading contributions as $\tau$ is varied, and where the Stokes lines are. To do so, we need the values of the bracketed potentials $[\Delta]_{\tau + r}$ for $r\in \bZ$. We find
\be
[\Delta]_{\tau + r} = \left[ \frac{2\tau -1}3 \right]_{\tau + r} = \begin{cases} \Delta + \dfrac{2r}3 &\text{if } r=0\mod 3 \\[.5em] \text{undefined} &\text{if } r = 1 \mod 3 \\[.5em] \Delta + \dfrac{2r-1}3 &\text{if } r = 2 \mod 3 \;. \end{cases}
\ee
In the second case the bracket is not defined because $\im \bigl( \Delta/(\tau + r) \bigr) \in\bZ\times\im \bigl( 1/(\tau + r) \bigr)$, \ie~because $\Delta$ sits exactly on the boundary of a strip defined by $\tau + r$. We can however consider $[\Delta]_{\tau+r}$ for values of $\Delta$ that are a bit off the boundary of the strip in the real direction. We consider the values $\Delta_{(\pm)} = \Delta \pm \epsilon$ with infinitesimal $\epsilon>0$ and find
\be
\bigl[ \Delta_{(+)} \bigr]_{\tau + r} \,\xrightarrow[\epsilon\to0]{}\, \Delta + \frac{2r-2}3 \;,\qquad \bigl[ \Delta_{(-)} \bigr]_{\tau + r} \,\xrightarrow[\epsilon\to0]{}\, \Delta + \frac{2r+1}3 \qquad\text{if } r = 1 \mod 3 \;.
\ee
Using these formulas, the values of $\Theta(\Delta, \tau+r)$ are easily computed.%
\footnote{For $r = 0 \mod 3$ one has to use the 1$^\text{st}$ case of $\Theta$, while for $r=2\mod 3$ the 2$^\text{nd}$ case.}
In particular, the imaginary parts of $\Theta$ computed on $\Delta_{(\pm)}$ are the same.

\begin{figure}[t]
\centering
\includegraphics[width=\textwidth]{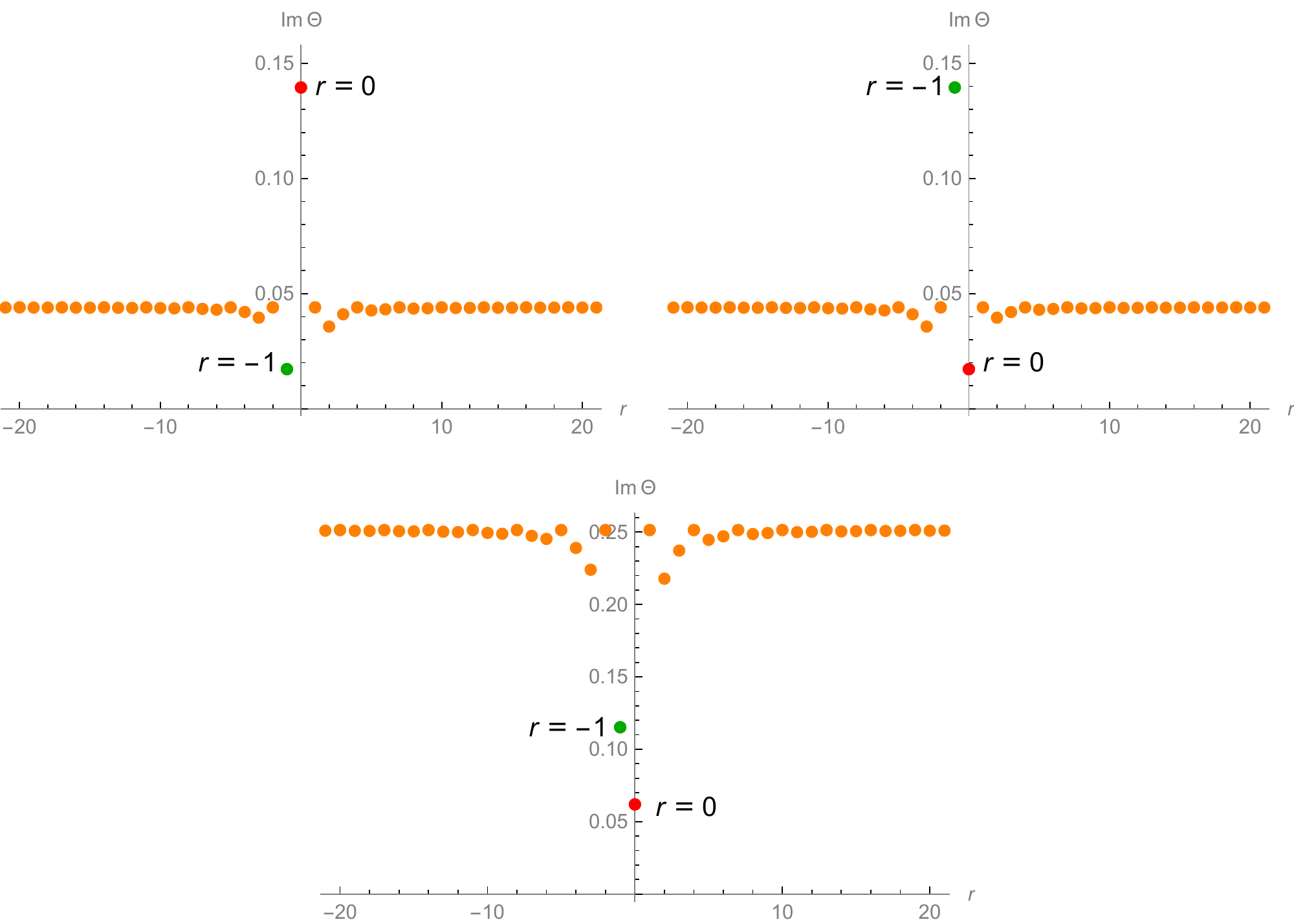}
\caption{The upper left plot shows the values of $\im\Theta(\Delta;\tau + r)$ as a function of $r$, for $\tau$ inside the semi-circle \eqref{red circle}. The red dot corresponds to $r=0$, and is the dominant contribution in this case. The upper right plot shows $\im\Theta(\Delta;\tau + r)$ for $\tau$ inside the semi-circle \eqref{green circle}. The green dot corresponds to $r=-1$, and is the dominant contribution in this case. The lower plot shows the values of $\im\Theta(\Delta;\tau + r)$ for $\tau$ outside the two semicircles, where there is no dominant contribution.
\label{fig: plots}}
\end{figure}

The dominant contribution to the index is determined by comparing the absolute values of $\exp\bigl( -\pi i N^2\Theta(\Delta;\tau + r) \bigr)$---or equivalently the imaginary parts of $\Theta$---as we vary $r$. When there is a particular value $\wh r$ for which $\im\Theta(\Delta; \tau + \wh r)$ is maximum, there is one dominant contribution which leads to a concrete estimate of the leading behavior of the index. When, instead, there is no maximum, we are left with an infinite number of competing contributions and more detailed information would be needed to resum them. We obtain the following values for the imaginary part of $\Theta$:
\be
\im\Theta(\Delta;\tau + r) = 
	\begin{cases}
		\ds \frac{2\im\tau}{27} \biggl(4 + \frac{\re\tau +r}{| \tau + r|^4} - \frac{3}{ |\tau + r|^2}\biggr) \quad& \text{if } r = 0 \mod 3 \\[1em]
		\ds \frac{8\im\tau}{27} & \text{if } r=1 \mod 3 \\[1em]
		\ds \frac{2\im\tau}{27}\biggl(4 - \frac{\re\tau +r}{|\tau + r|^4} - \frac{3}{|\tau + r|^2}\biggr)& \text{if } r = 2 \mod 3 \;.
	\end{cases}
\ee
Notice that the limiting value for large $|r|$ (equal to the value for $r=1\mod 3$) is as in (\ref{limiting value im Theta}). If there is a value of $r$ that maximizes $\im\Theta$, it must come from the first or third case. In particular, there exists $\wh r$ with $\wh r = 0 \mod 3$ if $\tau$ satisfies the following relation:
\be
\label{red circle}
\re\tau + \wh{r} > 3 \left\lvert \tau + \wh{r} \, \right\rvert^2 \qquad\text{ with }\qquad \wh r = 0 \mod 3 \;.
\ee
This corresponds to the interior of a semi-circle in the upper half $\tau$-plane, centered at the boundary point \mbox{$\tau = 1/6 -\wh{r}$} and with radius $1/6$. Similarly, there exists $\wh r$ with $\wh r = 2\mod 3$ if $\tau$ satisfies
\be
\label{green circle}
- \re\tau - \wh{r} > 3 \left\lvert \tau + \wh{r} \, \right\rvert^2 \qquad\text{ with }\qquad \wh r = 2 \mod 3 \;.
\ee
This corresponds to the interior of another semi-circle of radius $1/6$, centered at $\tau = -1/6 - \wh{r}$. The two inequalities (\ref{red circle}) and (\ref{green circle}) define two semi-circles in the fundamental range $0\leq \re\tau < 3$, for $\wh{r}=0$ and $\wh{r}=-1$ respectively, as well as all their images under the periodicity $\tau \to \tau+3$. On the other hand, outside the two regions there is no dominant contribution because, for all values of $r$, $\im\Theta$ is smaller than the limiting value. In Figure~\ref{fig: plots} we provide plots of $\im\Theta(\Delta;\tau+r)$ as $r$ is varied, both for $\tau$ insides the semi-circle (\ref{red circle}), inside the semi-circle (\ref{green circle}), and outside those two.

\begin{figure}[t]
\centering
\includegraphics[width=\textwidth]{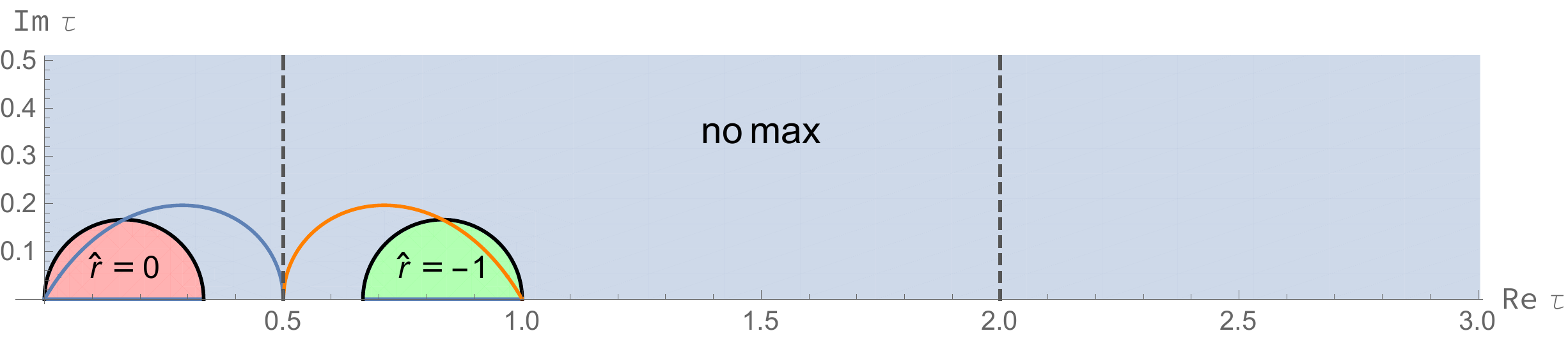}
\caption{Stokes lines in the $\tau$-plane. The red semicircle corresponds to the domain of analyticity where the $r=0$ contribution dominates the index. The green semicircle instead corresponds to the domain where the $r=-1$ contribution dominates. In the remaining region we do not have a dominant contribution. The blue and orange lines are the critical points of the entropy function for BPS black holes, in the two possible formulations. The dashed lines indicate the subspace where both fugacities $q$ and $y$ are real and the computation of \cite{Kinney:2005ej} applies.
\label{fig: tau}}
\end{figure}

\begin{figure}[t]
\centering
\includegraphics[scale=.7]{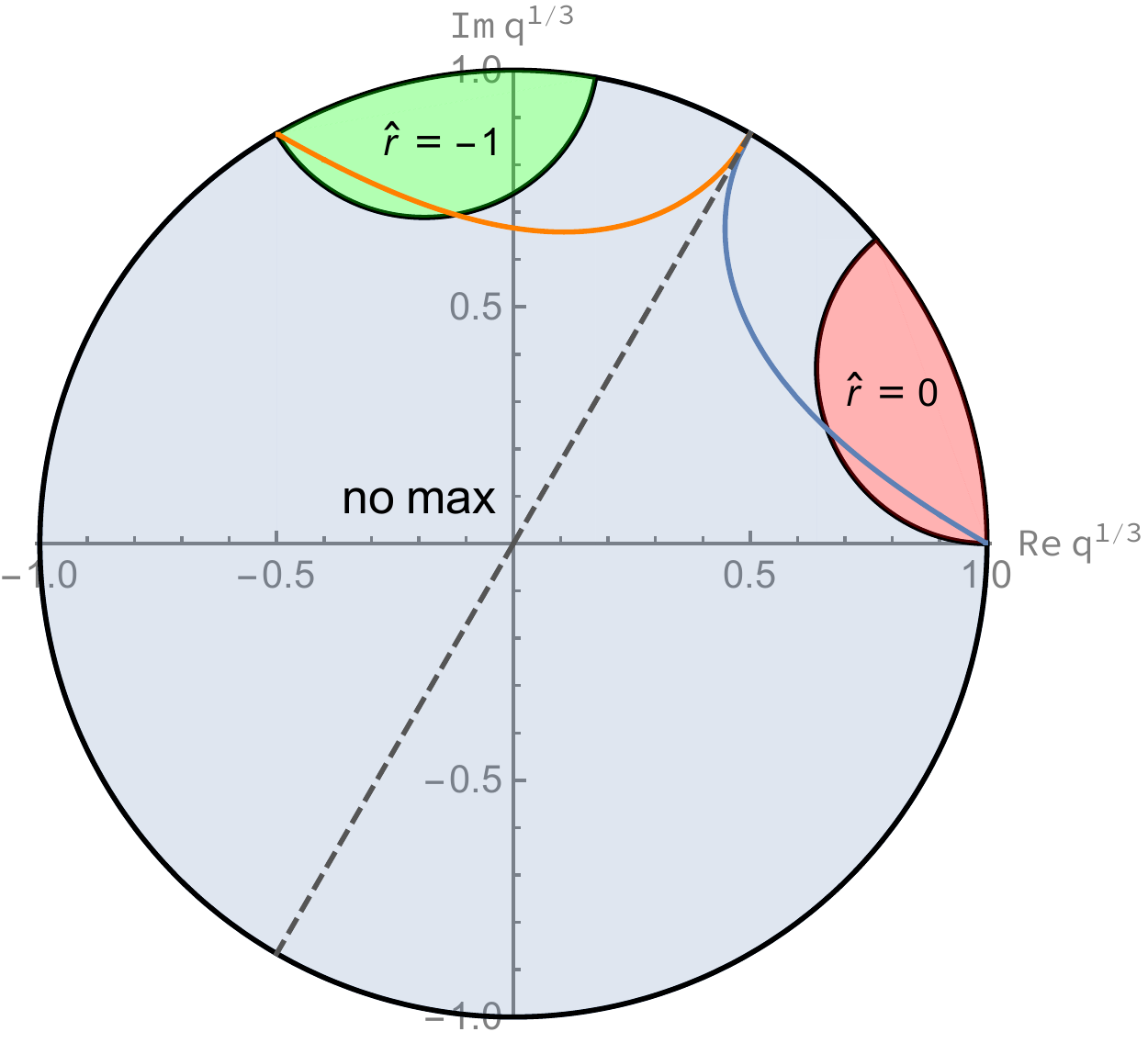}
\caption{Stokes lines in the $q$-plane, where the variable is $q^{1/3}$. The notation is the same as in Figure~\ref{fig: tau}.
\label{fig: q}}
\end{figure}

In Figure~\ref{fig: tau} we represent the fundamental range $0\leq \re\tau<3$ of the upper half $\tau$-plane, dividing it into regions according to the dominant contribution. In Figure~\ref{fig: q} we represent the same information in the $q$-plane, using $q^{1/3}$ as the variable. The red semi-circle \eqref{red circle} corresponds to the values of $\tau$ in which $\wh r = 0$, while the green semi-circle \eqref{green circle} corresponds to $\wh r=-1$. These are two different domains of analyticity. The remaining ``no max'' region, in blue, corresponds to values of $\tau$ for which there is no dominant contribution. The three regions are separated by Stokes lines (in black).

Inside the red semi-circle \eqref{red circle} the large $N$ limit of the superconformal index is
\be
\label{red index}
\log \, \cI_\infty(\Delta; \tau) = -\pi i N^2 \, \Theta(\Delta;\tau) = -\pi i N^2 \, \frac{[\Delta]_\tau^3}{\tau^2} = -\pi i N^2 \, \frac{(2\tau-1)^3}{27\tau^2} \;.
\ee
This expression exactly matches the entropy function (\ref{entropy function Zaffaroni}) of black holes with equal charges $Q$ and angular momenta $J$, and its Legendre transform selects the critical points (\ref{critical points Q J +1}). We represent the line of critical points, as $\mu>0$ is varied, by a blue solid line in Figures~\ref{fig: tau} and \ref{fig: q}. As we see from there, for $\mu>\mu_*$ the blue line lies inside the red semi-circle, meaning that the entropy of the single-center black hole is the dominant contribution to the index. This seems to confirm that ``large'' BPS black holes, with $Q > Q_*$ or equivalently $J > J_*$, are stable. On the contrary, for $0 < \mu < \mu_*$ the blue line plunges into the ``no max'' region. We can still identify the black hole entropy with the contribution of the basic solution (\ref{BA solution basic}) to the index, however such a contribution is no longer dominant. This suggests that ``small'' BPS black holes with $Q<Q_*$ might be unstable towards other supergravity configurations. We find the following values at the transition point:
\be
\mu_* = \frac23 \;,\qquad \tau_* = \frac{1+i}6 \;,\qquad Q_* = \frac89 \;,\qquad J_* = \frac{26}{27} \;,\qquad S_* = \frac{4\pi}3 \;,
\ee
where $Q,J,S$ are in units of $\nu$. It would be nice to derive these values from supergravity.

The green circle in Figures~\ref{fig: tau} and \ref{fig: q} corresponds to values of $\tau$ for which the $r=-1$ contribution dominates. In this domain we find
\bea
\label{green index}
\log \, \cI_\infty(\Delta;\tau) &= -\pi i N^2 \, \Theta(\Delta;\tau - 1) = -\pi i N^2 \, \Biggl( \frac{\bigl( [\Delta]_{\tau - 1} + 1 \bigr)^3}{(\tau - 1)^2} - 1 \Biggr) \\
&= -\pi i N^2 \, \left( \frac{(2\tau - 1)^3}{27(\tau - 1)^2} - 1\right) \;.
\eea
This also reproduces the entropy of single-center black holes: this expression matches the entropy function (\ref{entropy function Zaffaroni}) with the alternative constraint among the chemical potentials, given by (\ref{constraint on chem pot Zaffaroni}) with $+1$ substituted with $-1$. In the Figures we have indicated with a solid orange line the critical points obtained with the alternative extremization principle.

It is interesting to draw the subspace where both fugacities $q$ and $y$ are real and the computation of \cite{Kinney:2005ej} applies. We include this subspace both in Figure~\ref{fig: tau} and, in terms of $q^{1/3}$, in Figure~\ref{fig: q}. We see that the real subspace does not intercept the black hole lines: it only asymptotically reaches them, at the tail that describes black holes much smaller than the AdS radius.

\section*{Acknowledgements}
We thank A.~Arabi Ardehali, A.~Dabholkar, J.~Maldacena, S.~Minwalla, S.~Murthy and A.~Zaffaroni for useful discussions on this and related matters. F.B.~is supported in part by the MIUR-SIR grant RBSI1471GJ ``Quantum Field Theories at Strong Coupling: Exact Computations and Applications''.


\appendix


\section{Special functions}
\label{app: functions}

\paragraph{\matht{q}-Pochhammer symbol.} The function is defined as
\be
\label{def q-Poch}
(z;q)_\infty = \prod_{k=0}^\infty \bigl( 1 - zq^k \bigr) \;.
\ee
Here and in the following we set $z = e^{2\pi i u}$, $q = e^{2\pi i \tau}$ and take $|q|<1$.

\paragraph{Function \matht{\theta_0}.} This function, also called $q$-theta function, is defined as
\be
\label{def theta0}
\theta_0(u;\tau) = (z;q)_\infty \, (q/z; q)_\infty \;.
\ee
It satisfies the following relations:
\bea
\label{theta0 periodicities}
\theta_0(u + n + m\tau; \tau) &= (-z)^{-m} \, q^{- \frac{m(m-1)}2} \, \theta_0(u;\tau) \\
\theta_0(u; \tau) &= \theta_0(\tau-u; \tau) = - z\, \theta_0(-u; \tau) \;.
\eea
The modular transformation is
\be
\label{theta0 modular property}
\theta_0\left( \frac u\tau; - \frac1\tau \right) = e^{i\pi \left( \frac{u^2}\tau - u + \frac u\tau + \frac\tau6 + \frac1{6\tau} - \frac12 \right)} \, \theta_0(u; \tau) \;.
\ee
To derive it, one can relate $\theta_0$ to the Dedekind $\eta$ and Jacobi $\theta_3$ functions:
\be
\theta_0(u; \tau) = q^{\frac1{24}} \, \frac{\theta_3\bigl( u - \frac\tau2 - \frac12 ;\tau \bigr)}{\eta(\tau)} = \frac1{(q;q)_\infty} \sum_{n=-\infty}^{+\infty} (-z)^n \, q^{n(n-1)/2} \;,
\ee
where
\be
\theta_3(u; \tau) = \sum_{n = -\infty}^{+\infty} z^n \, q^{\frac{n^2}2} = (q;q)_\infty \, (-zq^{1/2}; q)_\infty \, (-z^{-1}q^{1/2};q)_\infty \;.
\ee
The function $\theta_3$ is also called $\vartheta_{00}$ in the literature.

\paragraph{Function \matht{\theta}.} This is a modification of the function $\theta_0$, defined as
\be
\theta(u; \tau) = e^{-i\pi u + i \pi \tau/6} \; \theta_0(u;\tau) \;.
\ee
Its periodicity relations are
\bea
\theta(u + n + m\tau; \tau) &= (-1)^{n+m} \, z^{-m} \, q^{-m^2/2} \, \theta(u; \tau) \\
\theta(-u; \tau) &= - \theta(u; \tau) \;.
\eea
The modular transformation is
\be
\theta\left( \frac u\tau; - \frac1\tau \right) = -i\, e^{i \pi u^2/\tau} \, \theta(u;\tau) \;.
\ee

\paragraph{Function \matht{\psi}.} Following \cite{Felder:1999}, we define the function
\be
\label{def function psi}
\psi(t) = \exp\left[ t \log \bigl( 1 - e^{-2\pi i t} \bigr) - \frac1{2\pi i} \Li_2 \bigl( e^{-2\pi i t} \bigr) \right] \;.
\ee
Within $\im t<0$, the definition is analytic and single-valued. The branch of the logarithm is determined by the series expansion $\log(1-z) = - \sum_{k=0}^\infty z^k/k$, whereas $\Li_2(z) = \sum_{k=1}^\infty z^k/k^2$. One can show that the branch cut ambiguities of the logarithm and the dilogarithm, that appear for $\im t\geq 0$, cancel in the definition of $\psi(t)$. This means that the latter function can be analytically continued to the whole complex plane yielding a meromorphic function.

Two useful properties of $\psi(t)$ are:
\be
\label{properties psi}
\psi(t) \, \psi(-t) = e^{-\pi i (t^2 - 1/6)} \;,\qquad\qquad \psi(t+n) = \bigl( 1 - e^{-2\pi i t} \bigr)^n \psi(t) \qquad \forall\, n \in \bZ \;,
\ee
valid for any $t\in \bC$.

\paragraph{Function \matht{\wt\Gamma}.} Setting now $p = e^{2\pi i \tau}$, $q = e^{2\pi i \sigma}$, the elliptic gamma function \cite{Felder:1999} is
\be
\wt\Gamma(u; \tau, \sigma) = \Gamma\bigl( z = e^{2\pi i u} \,;\, p = e^{2\pi i \tau} \,,\, q = e^{2\pi i \sigma} \bigr) = \prod_{m,n=0}^\infty \frac{ 1 - p^{m+1} q^{n+1} z^{-1} }{ 1 - p^m q^n z} \;,
\ee
defined for $|p|, |q|<1$. The function $\Gamma(z; p,q)$ is meromorphic in $z$, with simple zeros at $z = p^{m+1} q^{n+1}$ and simple poles at $z = p^{-m} q^{-n}$, for $m,n \in \bZ_{\geq 0}$. A plethystic representation is
\be
\Gamma(z; p,q) = \exp\left[ \sum_{k=1}^\infty \frac1k \; \frac{ z^k - p^k q^k z^{-k} }{ (1-p^k)(1-q^k)} \right] \;,
\ee
which is convergent for $|pq| < |z| < 1$.

The basic periodicity relations are
\bea
\wt\Gamma(u+1; \tau, \sigma) &= \wt\Gamma(u; \tau, \sigma) = \wt\Gamma(u; \sigma, \tau) \\
\wt\Gamma(u + \tau; \tau, \sigma) &= \theta_0(u; \sigma)\, \wt\Gamma(u; \tau, \sigma) \\
\wt\Gamma(u + \sigma; \tau, \sigma) &= \theta_0(u; \tau)\, \wt\Gamma(u; \tau, \sigma) \;.
\eea
There is also an inversion formula:
\be
\wt\Gamma(u; \tau, \sigma) = \frac1{\wt\Gamma( \tau + \sigma - u; \tau, \sigma)} \;.
\ee
For $\tau, \sigma, \tau/\sigma, \tau+\sigma \in \bC \setminus \bR$ there exist modular transformations:
\be
\label{modular transformations Gamma}
\wt\Gamma(u; \tau,\sigma) = e^{-\pi i \cQ(u; \tau,\sigma)} \, \frac{\wt\Gamma\bigl( \frac u\sigma; \frac\tau\sigma, - \frac1\sigma \bigr) }{ \wt\Gamma\bigl( \frac{u-\sigma}\tau; - \frac1\tau, - \frac\sigma\tau \bigr)} = e^{-\pi i \cQ(u; \tau,\sigma)} \, \frac{ \wt\Gamma\bigl( \frac u\tau; - \frac1\tau, \frac\sigma\tau \bigr) }{ \wt\Gamma\bigl( \frac{u-\tau}\sigma; - \frac\tau\sigma, - \frac1\sigma \bigr) }
\ee
where $\cQ(u; \tau, \sigma)$, defined in (\ref{def Q}), is a cubic polynomial in $u$.

In the degenerate case $\tau=\sigma$ the identities (\ref{modular transformations Gamma}) do not apply. However, there exists an alternative version \cite{Felder:1999}:
\be
\label{degenerate transformation Gamma}
\wt\Gamma(u; \tau, \tau) = \frac{ e^{-\pi i \cQ(u; \tau, \tau)} }{ \theta_0 \bigl( \frac u\tau; - \frac1\tau \bigr) } \; \prod_{k=0}^\infty \frac{ \psi\left( \frac{k+1+u}\tau \right) }{ \psi\left( \frac{k - u}\tau \right) } \;,
\ee
valid for $u \in \bC \setminus \{\bZ + \tau\bZ\}$. The function $\psi(t)$ is defined in (\ref{def function psi}).


\section{Real fugacities}
\label{app: real fugacities}

In Section~\ref{sec: large N} we evaluated, in the large $N$ limit, the contribution of some of the solutions to the BAEs to the sum in (\ref{eq:BAindex}). In particular we found that all $T$-transformed solutions (\ref{BA solution T-transformed}) of the basic solution, parameterized by the integer $r$, contribute at the same order in $N$, and their contributions are the arguments of $\wt\max$ in the final formula (\ref{index large N final}):
$$
- \pi i N^2\, \Theta(\Delta_1, \Delta_2; \tau + r)
$$
in terms of $\Theta$ defined in (\ref{def Theta}) and with $r\in \bZ$.

Here we show that when we take the fugacities $q, y_1, y_2$ to be all real, we end up precisely on a Stokes line. More precisely, we show that all contributions for $r \in \bZ$ organize into pairs, except for those elements that already sit on a Stokes line determined by the discontinuity of one of the functions $[ \,\cdot\,]_\tau$. In each pair, the two contributions have equal real part and compete. We cannot compute the sum of the two terms, as this would require more accurate information about the subleading corrections. Yet, this makes our result compatible with the result of \cite{Kinney:2005ej}. There it was found that, for real fugacities, the index scales as $\cO(1)$ at large $N$, implying that all $\cO(N^2)$ contributions cancel out. This point was also stressed in \cite{Choi:2018hmj, Choi:2018vbz}.

Real fugacities corresponds to chemical potentials whose real part is either zero or $-1/2$ modulo $1$. We distinguish the various possibilities into two major cases: the case that $0<q<1$, corresponding to $\tau \in i\bR_{\geq 0}$, and the case that $-1<q<0$, corresponding to $\tau \in -\frac12 + i\bR_{\geq0}$. Each case is further divided into subcases, according to the number of positive flavor fugacities $y_{1,2}$.

\hiddensubsection{The case \matht{0<q<1}}

We start with the case of positive angular fugacity, $0<q<1$. We take $\tau \in i\bR_{\geq0}$ and write
\be
\tau = i t \;,\qquad\text{ with } t>0 \;.
\ee
We distinguish three different subcases, corresponding to $y_1, y_2$ being both positive, one positive and one negative, or both negative.

If one of the flavor fugacities---that we call $y$---is real positive, we set the corresponding chemical potential
\be
\Delta = i \delta \;,\qquad\text{ with } \delta \in \bR \;.
\ee
We immediately see that $[\Delta]_\tau$ is not defined, because the argument sits precisely along one of the lines of discontinuity. On the other hand, for $r>0$ and generic $\delta$, the functions $[\Delta]_{\tau\pm r}$ are well-defined and we would like to evaluate them. We can precisely determine their values by splitting the imaginary axis in the $\Delta$-plane into a series of intervals
\be
I_k = (k, k+1) \times \frac tr \qquad\text{ with } k \in \bZ \;.
\ee
Assuming that $\delta \in I_k$, we see that $\Delta$ can be brought inside the strip corresponding to $\tau+r$ by shifting it by $k$, and inside the strip corresponding to $\tau-r$ by shifting it by $-k-1$. In formulas:
\be
\label{pos case, [ ] for I_k}
[\Delta]_{\tau + r} = i\delta + k \;,\qquad\qquad [\Delta]_{\tau - r} = i\delta - k -1 \qquad\qquad\text{for } \delta \in I_k \;.
\ee
If $\delta$ is equal to an extremum of $I_k$, \ie{} if $\delta = k \, t/r$ for some $k\in\bZ$, then $[\Delta]_{\tau \pm r}$ are not defined.

On the other hand, if one of the flavor fugacities---that we keep calling $y$---is real negative, we set its chemical potential
\be
\Delta = -\frac12 + i \delta \;,\qquad\text{ with } \delta \in \bR \;.
\ee
For $r>0$ and generic $\delta$, both functions $[\Delta]_{\tau \pm r}$ are well-defined. As before, we can determine their values by splitting the imaginary axis in intervals. This time the intervals are
\be
\wt I_k = (2k-1, 2k+1) \times \frac t{2r} \qquad\text{ with } k \in \bZ \;.
\ee
We then find
\be
\label{pos case, [ ] for tilde I_k}
[\Delta]_{\tau + r} = i \delta + k - \frac12 \;,\qquad\qquad [\Delta]_{\tau-r} = i \delta - k - \frac12 \qquad\qquad\text{for } \delta \in \wt I_k \;.
\ee
If $\delta = (2k-1)t/2r$ for some $k\in \bZ$, then $[\Delta]_{\tau\pm r}$ are not defined.

We now proceed to applying these formulas to the three subcases.

\subsubsection{The subcase \matht{0<y_1,y_2} with \matht{0<q<1}}
\label{app: case 0<y1 y2 q}

We take both flavor fugacities $y_{1,2}$ to be positive. Correspondingly, we set purely imaginary chemical potentials:
\be
\Delta_a = i \delta_a \;,\qquad\text{ with } \delta_a\in\bR \text{ and } a=1,2 \;.
\ee
We immediately see that neither $[\Delta_1]_\tau$, $[\Delta_2]_\tau$ nor $[\Delta_1+\Delta_2]_\tau$ are defined because their arguments sit precisely along one of the lines of discontinuity. This means that the contribution $r=0$ is already along a Stokes line.

Let us now consider $r > 0$. For generic $\delta_a$, the functions $[\Delta_a]_{\tau \pm r}$ are well-defined. Precisely, for $\delta_a \in I_{k_a}$ the functions $[\Delta_a]_{\tau\pm r}$ are given by (\ref{pos case, [ ] for I_k}). Turning to $\Delta_1 + \Delta_2$, we have two possibilities:
\be
\label{case positive (+,+) first}
\delta_1 + \delta_2 \in I_{k_1 + k_2} \quad\Rightarrow\quad \begin{aligned}
[\Delta_1 + \Delta_2]_{\tau+r} &= [\Delta_1]_{\tau + r} {+} [\Delta_2]_{\tau + r} \hspace{1.3em} = i (\delta_1 + \delta_2) + k_1 + k_2 \\
[\Delta_1 + \Delta_2]_{\tau-r} &= [\Delta_1]_{\tau - r} {+} [\Delta_2]_{\tau - r} {+}1 = i (\delta_1 + \delta_2) - k_1 - k_2 -1
\end{aligned}
\ee
or
\be
\label{case positive (+,+) second}
\delta_1 + \delta_2 \in I_{k_1+k_2+1} \;\;\Rightarrow\;\; \begin{aligned}
[\Delta_1 + \Delta_2]_{\tau+r} &= [\Delta_1]_{\tau + r} {+} [\Delta_2]_{\tau + r} {+}1 = i (\delta_1 + \delta_2) + k_1 + k_2 + 1 \\
[\Delta_1 + \Delta_2]_{\tau-r} &= [\Delta_1]_{\tau - r} {+} [\Delta_2]_{\tau - r} \hspace{1.3em} = i (\delta_1 + \delta_2) - k_1 - k_2 - 2 \,,
\end{aligned}
\ee
whereas $[\Delta_1 + \Delta_2]_{\tau \pm r}$ are not defined if $\delta_1 + \delta_2 = n \, t/r$ with $n \in\bZ$.

In the first case, given by (\ref{case positive (+,+) first}), we compute
\begin{align}
\Theta(\Delta_1,\Delta_2;\tau + r) &= \frac{[\Delta_1]_{\tau + r}[\Delta_2]_{\tau + r} \bigl( 2(\tau + r)-1-[\Delta_1]_{\tau + r}-[\Delta_2]_{\tau + r} \bigr)}{(\tau + r)^2} \nn \\
&= \frac{(i\delta_1 + k_1)(i\delta_2 + k_2)(2r - 1 - k_1 - k_2 + i(2t - \delta_1 - \delta_2))}{(r + it)^2} \\
\Theta(\Delta_1,\Delta_2;\tau - r) &= \frac{ \bigl( [\Delta_1]_{\tau - r}+1 \bigr) \bigl( [\Delta_2]_{\tau - r}+1 \bigr) \bigl( 2(\tau - r)-1-[\Delta_1]_{\tau - r}-[\Delta_2]_{\tau - r} \bigr)}{(\tau - r)^2} - 1 \nn \\
&= \frac{(i\delta_1 - k_1)(i\delta_2 - k_2)(-2r + 1 + k_1 + k_2 + i(2t - \delta_1 - \delta_2))}{(- r + it)^2} - 1 \;. \nn
\end{align}
In the second case, given by (\ref{case positive (+,+) second}), we compute
\begin{align}
\Theta(\Delta_1,\Delta_2;\tau + r) &= \frac{ \bigl( [\Delta_1]_{\tau + r}+1 \bigr) \bigl( [\Delta_2]_{\tau + r}+1 \bigr) \bigl( 2(\tau + r)-1-[\Delta_1]_{\tau + r}-[\Delta_2]_{\tau + r} \bigr)}{(\tau + r)^2} - 1 \nn \\
&= \frac{(i\delta_1 + k_1 + 1)(i\delta_2 + k_2 + 1)(2r - 1 - k_1 - k_2 + i(2t - \delta_1 - \delta_2))}{(r + it)^2} - 1 \nn \\
\Theta(\Delta_1,\Delta_2;\tau - r) &= \frac{[\Delta_1]_{\tau - r}[\Delta_2]_{\tau - r} \bigl( 2(\tau - r)-1-[\Delta_1]_{\tau - r}-[\Delta_2]_{\tau - r} \bigr)}{(\tau - r)^2} \\
&= \frac{(i\delta_1 - k_1 - 1)(i\delta_2 - k_2 - 1)(-2r + 1 + k_1 + k_2 + i(2t - \delta_1 - \delta_2))}{(- r + it)^2} \;. \nn
\end{align}
From these expression we see that, in both cases,
\be
\Theta(\Delta_1,\Delta_2;\tau - r) = - \, \overline{\Theta(\Delta_1,\Delta_2;\tau + r)} \, - 1 \;.
\ee
This implies that
\be
\im \Theta(\Delta_1,\Delta_2;\tau + r) = \im \Theta(\Delta_1,\Delta_2;\tau - r)
\ee
and thus
\be
\left| e^{-\pi i N^2 \Theta(\Delta_1, \Delta_2; \tau+r)} \right| = \left| e^{-\pi i N^2 \Theta(\Delta_1, \Delta_2; \tau-r)} \right| \;,
\ee
yielding to a competition between the two terms for each $r>0$.

\subsubsection{The subcase \matht{y_1 < 0 < y_2} with \matht{0<q<1}}

We take one flavor fugacity to be positive and one negative, say $y_1 < 0 < y_2$ (recall that the index is symmetric in the two flavor fugacities). Correspondingly, we set
\be
\Delta_1 = - \frac12 + i \delta_1 \;,\qquad \Delta_2 = i \delta_2 \;,\qquad\text{ with } \delta_{1,2} \in\bR \;.
\ee
Similarly to the previous case, $[\Delta_2]_\tau$ is not defined and the contribution $r=0$ is already along a Stokes line.

For $r > 0$ and generic $\delta_a$, instead, both functions $[\Delta_{1,2}]_{\tau \pm r}$ are well-defined. Assuming $\delta_1 \in \wt I_{k_1}$ the functions $[\Delta_1]_{\tau\pm r}$ are given by (\ref{pos case, [ ] for tilde I_k}), and assuming $\delta_2 \in I_{k_2}$ the functions $[\Delta_2]_{\tau\pm r}$ are given by (\ref{pos case, [ ] for I_k}). Turning to $\Delta_1 + \Delta_2$ we have two possibilities:
\be
\label{case positive (+,-) first}
\delta_1 + \delta_2 \in \wt I_{k_1+k_2} \quad \Rightarrow \quad \begin{aligned}
[\Delta_1 + \Delta_2]_{\tau+r} &= [\Delta_1]_{\tau + r} {+} [\Delta_2]_{\tau + r} \hspace{1.3em} = i (\delta_1 + \delta_2) + k_1 + k_2 - \frac{1}{2} \\
[\Delta_1 + \Delta_2]_{\tau-r} &= [\Delta_1]_{\tau - r} {+} [\Delta_2]_{\tau - r} {+}1 = i (\delta_1 + \delta_2) - k_1 - k_2 - \frac{1}{2}
\end{aligned}
\ee
or
\be
\label{case positive (+,-) second}
\delta_1 + \delta_2 \in \wt I_{k_1+k_2+1} \;\;\Rightarrow\;\; \begin{aligned}
[\Delta_1 + \Delta_2]_{\tau+r} &= [\Delta_1]_{\tau + r} {+} [\Delta_2]_{\tau + r} {+}1 = i (\delta_1 + \delta_2) + k_1 + k_2 + \frac12 \\
[\Delta_1 + \Delta_2]_{\tau-r} &= [\Delta_1]_{\tau - r} {+} [\Delta_2]_{\tau - r} \hspace{1.3em} = i (\delta_1 + \delta_2) - k_1 - k_2 - \frac32 \,,
\end{aligned}
\ee
whereas $[\Delta_1 + \Delta_2]_{\tau \pm r}$ are not defined if $\delta_1 + \delta_2 = (2n - 1) \, t/2r$ with $n \in\bZ$.

In the first case, given by (\ref{case positive (+,-) first}), we compute
\begin{align}
\Theta(\Delta_1,\Delta_2;\tau + r) &= \frac{(i\delta_1 + k_1 - \frac12)(i\delta_2 + k_2)(2r - \frac12 - k_1 - k_2 + i(2t - \delta_1 - \delta_2))}{(r + it)^2} \\
\Theta(\Delta_1,\Delta_2;\tau - r) &= \frac{(i\delta_1 - k_1+ \frac12)(i\delta_2 - k_2)(-2r + \frac12 + k_1 + k_2 + i(2t - \delta_1 - \delta_2))}{(- r + it)^2} -1 \;. \nn
\end{align}
In the second case, given by (\ref{case positive (+,-) second}), we compute
\bea
\Theta(\Delta_1,\Delta_2;\tau + r) &= \frac{(i\delta_1 + k_1 + \frac12)(i\delta_2 + k_2 + 1)(2r - \frac12 - k_1 - k_2 + i(2t - \delta_1 - \delta_2))}{(r + it)^2} -1 \\
\Theta(\Delta_1,\Delta_2;\tau - r) &= \frac{(i\delta_1 - k_1 - \frac12)(i\delta_2 - k_2 - 1)(-2r + \frac12 + k_1 + k_2 + i(2t - \delta_1 - \delta_2))}{(- r + it)^2} \;.
\eea
Hence, in both cases we find $\Theta(\Delta_1,\Delta_2;\tau - r) = - \, \overline{\Theta(\Delta_1,\Delta_2;\tau + r)} \, - 1$, meaning that the two terms compete.

\subsubsection{The subcase \matht{y_1, y_2 < 0} with \matht{0<q<1}}

We take both flavor fugacities $y_{1,2}$ to be negative. Correspondingly we set
\be
\Delta_a = - \frac{1}{2} + i \delta_a \;,\qquad \text{ with } \delta_a \in\bR \text{ and } a=1,2 \;.
\ee
In this case $[\Delta_a]_\tau$ are defined but $[\Delta_1 + \Delta_2]_\tau$ is not. Therefore the contribution $r=0$ is along a Stokes line.

For $r > 0$ and generic $\delta_a$, the functions $[\Delta_a]_{\tau \pm r}$ are well-defined. Assuming $\delta_a \in \wt I_{k_a}$ then $[\Delta_a]_{\tau\pm r}$ are given by (\ref{pos case, [ ] for tilde I_k}). We have two possibilities for $\Delta_1 + \Delta_2$:
\be
\label{case positive (-,-) first}
\delta_1 + \delta_2 \in I_{k_1+k_2 - 1} \;\;\Rightarrow\;\; \begin{aligned}
[\Delta_1 + \Delta_2]_{\tau+r} &= [\Delta_1]_{\tau + r} {+} [\Delta_2]_{\tau + r} \hspace{1.3em} = i (\delta_1 + \delta_2) + k_1 + k_2 - 1 \\
[\Delta_1 + \Delta_2]_{\tau-r} &= [\Delta_1]_{\tau - r} {+} [\Delta_2]_{\tau - r} {+} 1 = i (\delta_1 + \delta_2) - k_1 - k_2
\end{aligned}
\ee
or
\be
\label{case positive (-,-) second}
\delta_1 + \delta_2 \in I_{k_1+k_2} \quad\Rightarrow\quad \begin{aligned}
[\Delta_1 + \Delta_2]_{\tau+r} &= [\Delta_1]_{\tau + r} {+} [\Delta_2]_{\tau + r} {+} 1 = i (\delta_1 + \delta_2) + k_1 + k_2 \\
[\Delta_1 + \Delta_2]_{\tau-r} &= [\Delta_1]_{\tau - r} {+} [\Delta_2]_{\tau - r} \hspace{1.3em} = i (\delta_1 + \delta_2) - k_1 - k_2 - 1,
\end{aligned}
\ee
whereas $[\Delta_1 + \Delta_2]_{\tau \pm r}$ are not defined if $\delta_1 + \delta_2 = n \, t/r$ with $n\in\bZ$. In the first case, given by (\ref{case positive (-,-) first}), we compute
\be
\Theta(\Delta_1,\Delta_2;\tau + r) = \frac{(i\delta_1 + k_1 - \frac12)(i\delta_2 + k_2 - \frac12)(2r - k_1 - k_2 + i(2t - \delta_1 - \delta_2))}{(r + it)^2} \;,
\ee
while in the second case, given by (\ref{case positive (-,-) second}), we compute
\be
\Theta(\Delta_1,\Delta_2;\tau + r) = \frac{(i\delta_1 + k_1 + \frac12)(i\delta_2 + k_2 + \frac12)(2r - k_1 - k_2 + i(2t - \delta_1 - \delta_2))}{(r + it)^2} -1 \;,
\ee
and in both cases $\Theta(\Delta_1,\Delta_2;\tau - r) = - \, \overline{\Theta(\Delta_1,\Delta_2;\tau + r)} \, -1$.

\hiddensubsection{The case \matht{-1<q<0}}

Now we move to the case of negative angular fugacity, $-1< q<0$, and set
\be
\tau = -\frac12 + i t \;,\qquad\text{ with } t>0 \;.
\ee
Once again, we distinguish three different subcases corresponding to $y_1, y_2$ being both positive, one positive and one negative, or both negative. First, let us discuss the new intervals we need.

If a flavor fugacity $y$ is real positive, as before we set $\Delta = i\delta$ with $\delta \in \bR$. Taking $r\geq 0$ and generic $\delta$, the functions $[\Delta]_{\tau + r +1}$ and $[\Delta]_{\tau - r}$ are well-defined. To evaluate them, we split the imaginary axis into intervals
\be
I_k = (k, k+1) \times \frac{2t}{2r+1} \qquad\text{ with } k \in \bZ \;.
\ee
We find
\be
\label{neg case, [ ] for I_k}
[\Delta]_{\tau + r + 1} = i\delta + k \;,\qquad\qquad [\Delta]_{\tau - r} = i\delta - k -1 \qquad\qquad\text{for } \delta \in I_k \;.
\ee
If $\delta = n \, 2t/(2r + 1)$ for some $n\in\bZ$, then $[\Delta]_{\tau + r + 1}$ and $[\Delta]_{\tau - r}$ are not defined.

On the other hand, if a flavor fugacity $y$ is real negative, we set $\Delta = -\frac12 + i \delta$ with $\delta \in \bR$. For $r\geq 0$ and generic $\delta$, the functions $[\Delta]_{\tau + r +1}$ and $[\Delta]_{\tau - r}$ are once again well-defined. We split the imaginary axis into intervals
\be
\wh I_k = ( 2k- 1, 2k +1 ) \times \frac t{2r + 1} \qquad\text{ with } k \in \bZ \;.
\ee
This time we find
\be
\label{neg case, [ ] for hat I_k}
[\Delta]_{\tau + r + 1} = i\delta + k - \frac12 \;,\qquad\qquad [\Delta]_{\tau - r} = i\delta - k - \frac12 \qquad\qquad\text{for } \delta \in \wh I_k \;.
\ee
If $\delta = (2n_1 - 1) \, t/(2r + 1)$ for some $n\in\bZ$, then $[\Delta]_{\tau + r + 1}$ and $[\Delta]_{\tau - r}$ are not defined.

\subsubsection{The subcase \matht{0<y_1, y_2} with \matht{-1<q<0}}
\label{app: case q < 0 < y1 y2}

We take both flavor fugacities $y_{1,2}$ to be positive and set
\be
\Delta_a = i \delta_a \;,\qquad\text{ with } \delta_a \in\bR \text{ and } a=1,2 \;.
\ee
For $r\geq 0$ and generic $\delta_a \in I_k$, the functions $[\Delta_a]_{\tau + r + 1}$ and $[\Delta_a]_{\tau - r}$ are well-defined and given by (\ref{neg case, [ ] for I_k}). There are then two possibilities. If $\delta_1 + \delta_2 \in I_{k_1+k_2}$ then
\bea {}
[\Delta_1 + \Delta_2]_{\tau+r+1} &= [\Delta_1]_{\tau + r + 1} + [\Delta_2]_{\tau + r + 1} = i (\delta_1 + \delta_2) + k_1 + k_2 \\
[\Delta_1 + \Delta_2]_{\tau-r} &= [\Delta_1]_{\tau - r} + [\Delta_2]_{\tau - r} + 1 \hspace{.15em} = i (\delta_1 + \delta_2) - k_1 - k_2 -1
\eea
and
\be
\Theta(\Delta_1,\Delta_2;\tau + r + 1) = \frac{(i\delta_1 + k_1)(i\delta_2 + k_2)(2r - k_1 - k_2 + i(2t - \delta_1 - \delta_2))}{(r + \frac12 + it)^2} \;.
\ee
If $\delta_1 + \delta_2 \in I_{k_1+k_2+1}$ then
\bea {}
[\Delta_1 + \Delta_2]_{\tau+r+1} &= [\Delta_1]_{\tau + r + 1} + [\Delta_2]_{\tau + r + 1} +1 = i (\delta_1 + \delta_2) + k_1 + k_2 + 1 \\
[\Delta_1 + \Delta_2]_{\tau-r} &= [\Delta_1]_{\tau - r} + [\Delta_2]_{\tau - r} \hspace{3.6em} = i (\delta_1 + \delta_2) - k_1 - k_2 - 2
\eea
and
\be
\Theta(\Delta_1,\Delta_2;\tau + r + 1) = \frac{(i\delta_1 + k_1 + 1)(i\delta_2 + k_2 + 1)(2r - k_1 - k_2 + i(2t - \delta_1 - \delta_2))}{(r - \frac12 + it)^2} -1 \;.
\ee
If $\delta_1 + \delta_2 = n \,2t/(2r + 1)$ with $n\in\bZ$, then $[\Delta_1 + \Delta_2]_{\tau + r + 1}$ and $[\Delta_1 + \Delta_2]_{\tau - r}$ are not defined.

In both well-defined cases, we find
\be
\Theta(\Delta_1,\Delta_2;\tau - r) = - \, \overline{\Theta(\Delta_1,\Delta_2;\tau + r + 1)} \, -1
\ee
This implies that
\be
\left| e^{-\pi i N^2 \Theta(\Delta_1,\Delta_2;\tau + r + 1)} \right| = \left| e^{-\pi i \Theta(\Delta_1,\Delta_2;\tau - r)} \right| \;,
\ee
yielding to a competition between the two terms for each $r\geq 0$.

\subsubsection{The subcase \matht{y_1 < 0 < y_2} with \matht{-1<q<0}}

We take one flavor fugacities to be positive and the other one to be negative. Hence we set
\be
\Delta_1 = - \frac{1}{2} + i \delta_1 \;,\qquad \Delta_2 =  i \delta_2 \;,\qquad\text{ with } \delta_{1,2} \in\bR \;.
\ee
For $r \geq 0$ and generic $\delta_a$, the functions $[\Delta_a]_{\tau + r + 1}$ and $[\Delta_a]_{\tau - r}$ are well-defined. Assuming $\delta_1 \in \wh I_{k_1}$ and $\delta_2 \in I_{k_2}$, those functions are given by (\ref{neg case, [ ] for hat I_k}) and (\ref{neg case, [ ] for I_k}), respectively. If $\delta_1 + \delta_2 \in \wh I_{k_1+k_2}$ then
\bea {}
[\Delta_1 + \Delta_2]_{\tau+r+1} &= [\Delta_1]_{\tau + r + 1} + [\Delta_2]_{\tau + r + 1} = i (\delta_1 + \delta_2) + k_1 + k_2 - \frac12  \\
[\Delta_1 + \Delta_2]_{\tau-r} &= [\Delta_1]_{\tau - r} + [\Delta_2]_{\tau - r} +1 \hspace{.15em} = i (\delta_1 + \delta_2) - k_1 - k_2 - \frac{1}{2}
\eea
and
\be
\Theta(\Delta_1,\Delta_2;\tau + r + 1) = \frac{(i\delta_1 + k_1 - \frac12)(i\delta_2 + k_2)(2r + \frac12 - k_1 - k_2 + i(2t - \delta_1 - \delta_2))}{(r + \frac12 + it)^2} \;.
\ee
If $\delta_1 + \delta_2 \in \wh I_{k_1+k_2+1}$ then
\bea {}
[\Delta_1 + \Delta_2]_{\tau+r+1} &= [\Delta_1]_{\tau + r + 1} + [\Delta_2]_{\tau + r + 1} + 1 = i (\delta_1 + \delta_2) + k_1 + k_2 + \frac12 \\
[\Delta_1 + \Delta_2]_{\tau-r} &= [\Delta_1]_{\tau - r} + [\Delta_2]_{\tau - r} \hspace{3.6em} = i (\delta_1 + \delta_2) - k_1 - k_2 - \frac32
\eea
and
\be
\Theta(\Delta_1,\Delta_2;\tau + r + 1) = \frac{(i\delta_1 + k_1 + \frac12)(i\delta_2 + k_2 + 1)(2r + \frac12 - k_1 - k_2 + i(2t - \delta_1 - \delta_2))}{(r - \frac12 + it)^2} -1 \,.
\ee
If $\delta_1 + \delta_2 = (2n - 1) t/(2r + 1)$ with $n\in\bZ$, then $[\Delta_1 + \Delta_2]_{\tau + r + 1}$ and $[\Delta_1 + \Delta_2]_{\tau - r}$ are not defined. In both well-defined cases we find $\Theta(\Delta_1,\Delta_2;\tau - r) = - \, \overline{\Theta(\Delta_1,\Delta_2;\tau + r + 1)}-1$.

\subsubsection{The subcase \matht{y_1, y_2 <0} with \matht{-1 < q<0}}

Finally, we consider both flavor fugacities to be negative and set
\be
\Delta_a = - \frac12 + i \delta_a \;,\qquad\text{ with } \delta_a\in\bR \text{ and } a=1,2 \;.
\ee
For $r\geq0$ and generic $\delta_a \in \wh I_{k_a}$, the functions $[\Delta_a]_{\tau + r + 1}$ and $[\Delta_a]_{\tau - r}$ are well-defined and given by (\ref{neg case, [ ] for hat I_k}). If $\delta_1 + \delta_2 \in I_{k_1+k_2 - 1}$ then
\bea {}
[\Delta_1 + \Delta_2]_{\tau+r+1} &= [\Delta_1]_{\tau + r + 1} + [\Delta_2]_{\tau + r + 1} = i (\delta_1 + \delta_2) + k_1 + k_2 - 1 \\
[\Delta_1 + \Delta_2]_{\tau-r} &= [\Delta_1]_{\tau - r} + [\Delta_2]_{\tau - r} +1 \hspace{.15em} = i (\delta_1 + \delta_2) - k_1 - k_2
\eea
and
\be
\Theta(\Delta_1,\Delta_2;\tau + r + 1) = \frac{(i\delta_1 + k_1 - \frac12)(i\delta_2 + k_2 - \frac12)(2r + 1 - k_1 - k_2 + i(2t - \delta_1 - \delta_2))}{(r + \frac12 + it)^2} \;.
\ee
If $\delta_1 + \delta_2 \in I_{k_1+k_2+1}$ then
\bea {}
[\Delta_1 + \Delta_2]_{\tau+r+1} &= [\Delta_1]_{\tau + r + 1} + [\Delta_2]_{\tau + r + 1} + 1 = i (\delta_1 + \delta_2) + k_1 + k_2  \\
[\Delta_1 + \Delta_2]_{\tau-r} &= [\Delta_1]_{\tau - r} + [\Delta_2]_{\tau - r} \hspace{3.6em} = i (\delta_1 + \delta_2) - k_1 - k_2 - 1
\eea
and
\be
\Theta(\Delta_1,\Delta_2;\tau + r + 1) = \frac{(i\delta_1 + k_1 + \frac12)(i\delta_2 + k_2 + \frac12)(2r + 1 - k_1 - k_2 + i(2t - \delta_1 - \delta_2))}{(r - \frac12 + it)^2} -1 \,.
\ee
If $\delta_1 + \delta_2 = n \,2t/(2r + 1)$ with $n\in\bZ$, then $[\Delta_1 + \Delta_2]_{\tau + r + 1}$ and $[\Delta_1 + \Delta_2]_{\tau - r}$ are not defined. In both well-defined cases: $\Theta(\Delta_1,\Delta_2;\tau - r) = - \, \overline{\Theta(\Delta_1,\Delta_2;\tau + r + 1)} -1$.


\bibliographystyle{ytphys}
\bibliography{BHentropy}
\end{document}